# Long-Term Cyclicities in Phanerozoic Sea-Level Sedimentary Record and their Potential Drivers

Slah Boulila[a,b*], Jacques Laskar[b], Bilal U. Haq[a,c], Bruno Galbrun[a], Nathan Hara[b]

[a]*Sorbonne Universités, UPMC Univ Paris 06, CNRS, Institut des Sciences de la Terre de Paris (ISTeP), 4 place Jussieu 75005 Paris, France*

[b]*Astronomie et Systèmes Dynamiques IMCCE, Observatoire de Paris, 77 Avenue Denfert-Rochereau, 75014 Paris, France*

[c]*Smithsonian Institution, Washington DC, USA*

[*] Corresponding author.

*E-mail address*: slah.boulila@upmc.fr

**Supplementary material**

**SM-1**. Supplementary material on (1) time-series analysis, (2) very long-term eustatic cycles vs. other geological process during the Phanerozoic Eon, and (3) Phanerozoic $\delta^{18}O$ data.

**SM-2**. Supplementary material on galactic trajectories and cosmic rays.

ABSTRACT

Cyclic sedimentation has varied at several timescales and this variability has been geologically well documented at Milankovitch timescales, controlled in part by climatically (insolation) driven sea-level changes.

At the longer (tens of Myr) timescales connection between astronomical parameters and sedimentation via cyclic solar-system motions within the Milky Way has also been proposed, but this hypothesis remains controversial because of the lack of long geological records. In addition, the absence of a meaningful physical mechanism that could explain the connection between climate and astronomy at these longer timescales led to the more plausible explanation of plate motions as the main driver of climate and sedimentation through changes in ocean and continent mass distribution on Earth.

Here we statistically show a prominent and persistent ~36 Myr sedimentary cyclicity superimposed on two megacycles (~250 Myr) in a relatively well-constrained sea-level (SL) record of the past 542 Myr (Phanerozoic eon). We also show two other significant ~9.3 and ~91 Myr periodicities, but with lower amplitudes. The ~9.3 Myr cyclicity was previously attributed to long-period Milankovitch band based on the Cenozoic record. However, the ~91 Myr cyclicity has never been observed before in the geologic record. The ~250 Myr cyclicity was attributed to the Wilson tectonic (supercontinent) cycle. The ~36 Myr periodicity, also detected for the first time in SL record, has previously been ascribed either to tectonics or to astronomical cyclicity.

Given the possible link between amplitudes of the ~36 and ~250 Myr cyclicities in SL record and the potential that these periodicities fall into the frequency band of solar system motions, we suggest an astronomical origin, and model these periodicities as originating from the path of the solar system in the Milky Way as vertical and radial periods that modulate the flux of cosmic rays on Earth. Our finding of the ~36 Myr SL cyclicity lends credibility to the existing hypothesis about the imprint of solar-system vertical period on the geological record. The ~250 Myr megacycles are tentatively attributed to a radial period. However, tectonic causal mechanisms remain equally plausible.

The potential existence of a correlation between the modeled astronomical signal and the geological record may offer an indirect proxy to understand the structure and history of the Milky Way by providing a 542 Myr long record of the path of the Sun in our Galaxy.

## 1. Introduction

Increasing evidence from high-resolution sediment records from ocean drilling programs (Zachos et al., 2001; Cramer et al., 2009; Friedrich *et al.*, 2012) and a parallel development of well-constrained astronomical models (Laskar et al., 2004, 2011) suggests that significant astro-climatic (Milankovitch, 1941) variability is present at million year (Myr) to multi-Myr timescales (e.g., Pälike et al., 2006; Boulila et al., 2011, 2012). The study of the influence of Myr to multi-Myr Milankovitch cycles to climate change and sedimentation is of considerable value in deciphering biological turnover (e.g., van Dam et al., 2006), carbon-cycle variations (Pälike et al., 2006, Boulila et al., 2012), ice-sheet events (Zachos et al., 2001; Pälike et al., 2006), sea-level fluctuations (Boulila et al., 2011), etc. Cyclicities of tens of Myr to few hundreds of Myr have also been detected in the geological records (e.g., Raup and Sepkoski, 1984; Rohde and Muller, 2005; Svensmark, 2006; Meyers and Peters, 2011), but the causes of these cyclicities are not well understood. Two potential drivers have been proposed for these very long geological periodicities: 1) major plate tectonic motions (e.g., Cogné and Humler, 2006; DeCelles et al., 2009; Meyers and Peters, 2011), and 2) the motions of the Solar System in the Milky Way (e.g., Svensmark, 2006; Medvedev and Melott, 2007).

The effect of tectonics on climate and sedimentation has been hypothesized through changing ocean volume as well as the geographic re-arrangement of continental and ocean mass distribution (e.g., Vail et al., 1977; Haq et al., 1987; Lagabrielle et al., 2009; Meyers and Peters, 2011; Cloetingh and Haq 2015 ; Zaffos et al., 2017). Changes in galactic cosmic-ray (GCR) flux induced by periodic motions of the Solar System in the Galaxy have been also been proposed to explain the link between climate and astronomy at tens to hundreds of Myr (e.g., Shaviv, 2002).

Since 1980's several researchers have focused on the imprint of Solar System motions in the Millky Way in the geological record, because of the potential implications for Earth's climate change (Shaviv and Veizer, 2003) and biotic mass extinctions (Raup and Sepkoski, 1984; Medvedev and Melott, 2007; Svensmark, 2012). However, previous studies were hampered by the low resolution of geological records, or by the lack of understanding of the significance of the geological variations in terms of climate or environmental factors. For instance, biodiversity variations were used to relate environmentally-driven cyclic extinctions to cometary impacts on Earth (Raup and Sepkoski, 1984) or to Earth's climate change induced by the galactic cosmic ray (GCR) flux (Medvedev and Melott, 2007; Svensmark, 2012), both in response to Solar System motions in the Milky Way.

The hypothesis of the influence of GCR on climate is being debated because of the absence of an arguable physical mechanism that can explain the connection between GCR and temperature. The most popular suggested mechanism is the formation of cloud layer that modifies Earth's albedo (Shaviv, 2002; Shaviv and Veizer, 2003; Svensmark, 2006, 2007). Although it has not been suggested to be the only effective parameter, GCR surprisingly seem to have a significant impact on Earth's climate, evidenced through the remarkable correlation between cosmic ray and climate variations (Svensmark, 2007).

Despite these constraints, there is some consensus that the Solar System vertically oscillates across the Galactic midplane, with a half-period ranging from 26 to 41 Myr (Bahcall and Bahcall, 1985; Stothers, 1998; Randal and Reece, 2014). In addition, it has been proposed that the Solar System moves around the galactic center with an orbital period of ~240 Myr (Svensmark, 2006), with a radial period of 180 Myr (Bailer-Jones, 2009). But given our limited knowledge of gravitational potential of the Galaxy, these two periods can in fact range over a large set of values, with the only observational constraint being the measured present velocity of the Sun, with respect to an inertial frame, as proposed by Schönrich *et al.* (2010) (Supplementary material SM-2).

Here we statistically quantify long-term cyclic sea-level (SL) variations from a Phanerozoic SL data that provide a rare longer-term and continuous record of geological parameters. We show that some detected SL periodicities are of the same order as those predicted for the Solar System motions in the Galaxy. Then, we astronomically model these dominant geological periodicities by fitting them to galactic parameters, with the aim to provide constraints on the gravitational potential within the Milky Way. Finally, we discuss other possible causal mechanisms of SL change at these timescales.

## 2. Data and methods

The eustatic sea level curve for the Phanerozoic is constructed based on sequence-stratigraphic studies of the marine sedimentary sections around the world. The details of these methodologies are described in Haq et al. (1987) for the Meso-Cenozoic and in Haq and Schutter (2008) for the Paleozoic. Sections for such studies mostly come from the ancient continental margins or interior basins where base-level fluctuations of the ocean (advance and retreat of shoreline) are best preserved. Where possible, margins with tectonic quiescence (or where tectonic influences can be corrected for) are chosen (such as peri-cr areas), to be able to isolate the eustatic signal from the tectonic (local subsidence or uplift) one. Correlations are based on the available marine biostratigraphic markers, and if an event can be demonstrated to be present in a number of non-contiguous sections, they are considered wide-spread and therefore eustatic.

We have compiled the Ceno-Mesozoic eustatic curve (Haq *et al.*, 1987), revised through a comparison with the Arabian Platform records (Haq and Al-Qahtani, 2005), and the most recent Paleozoic eustatic curve (Haq and Schutter, 2008). This unified Phanerozoic (0 to 542 Ma) sea level curve (Fig. 1) allows a regular temporal resolution of 0.1 Myr and has been recalibrated to recently updated geological timescale (Ogg *et al.*, 2016).

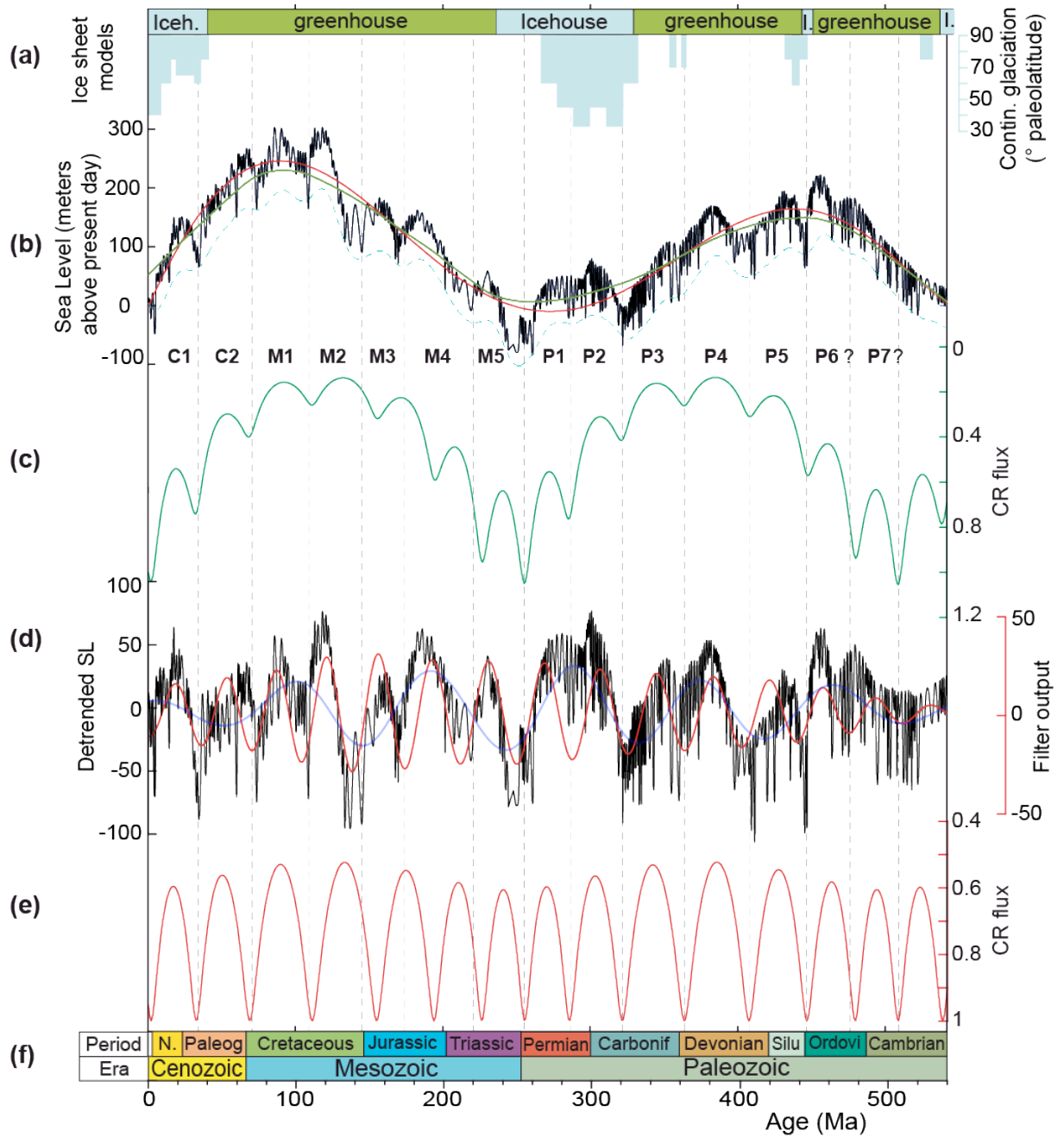

**Fig. 1** Comparison of Phanerozoic eustatic curve, ice sheet models, and solar-system motion induced cosmic-ray (CR) models. **(a)** Ice sheet models, upper model (greenhouse in white boxes, and icehouse in grey boxes) from (Zachos *et al.*, 2008; Frakes *et al.*, 1992),

lower model (in vertical grey bars for icehouse intervals) from (Ridgwell, 2005). **(b)** Original Phanerozoic eustatic curve (after Haq *et al.,* 1987, Haq and Al-Qahtani, 2005, and Haq and Schutter, 2008). Roughly Meso-Cenozoic and Paleozoic megacycles are fitted by 25% weighted average (dashed curve) and six-order polynomial (solid curve) methods (see Fig. 3). Gaussian lowpass filter output (0 to 0.05 $Myr^{-1}$ frequency cutoff) performed to uncover both ~36 and ~250 Myr cycles is also shown (dashed curve); C1, C2, M1-M5, and P1-P7 are the visually inspected ~36 Myr cycles (roughly 'C' for Cenozoic, 'M' for Mesozoic, and 'P' for Paleozoic), question marks indicate that ~36 Myr cycle boundaries are uncertain. **(c)** CR astronomical modeling (inverted axis) showing the ~36 and ~250 Myr vertical and radial motions of the solar system in the Milky Way (with a decay in $1/d^2$, see Supplementary material SM-2); times of midplane crossings and galactic center nearing correspond to periods of higher CR, climatic cooling, ice-sheet formation and glacioeustatic sea-level falls (see text for discussion). **(d)** Detrended eustatic curve (residuals of the 25% weighted average shown in 'b') and Gaussian bandpass filter outputs: 0.028 ± 0.012 $Myr^{-1}$ frequency cutoff to recover the ~36 Myr cyclicity, and 0.0103 ± 0.005 $Myr^{-1}$ frequency cutoff to recover the ~91 Myr cyclicity. **(e)** CR astronomical modeling (inverted axis) showing the ~36 Myr vertical motion of the solar system in the galaxy (with a decay in $1/d^2$, see Supplementary material SM-2). **(f)** Geologic Time Scale (after Ogg *et al.*, 2016).
"

To quantify the long and short-term cyclicities, we performed spectral analysis using the multitaper method (MTM) associated with the robust noise modeling (Ghil *et al.*, 2002) as implemented in the 'astrochron' freeware (Meyers, 2014) (Fig. 2). For false "significant" spectral peaks of multiple null-hypothesis testing, especially for statistical significance of the dominant 36 Myr period, we have used the Bonferroni correction (Fig. 2). The megacycles were measured using both weighted average and polynomial methods, then subtracted to highlight the shorter-term cycles (Figs. 1 and 2). The effect of detrending on the spectral quantification of shorter periods has been extensively tested. Here we automatically compared a pure period retrieved from the undetrended original SL data using the Generalized Lomb-Scargle (GLS) periodogram (Zechmeister and Kürster, 2009) with periods estimated (with the same GLS method) after successive polynomial-order detrends (Fig. 3).

To study the continuity of cycles throughout the Phanerozoic, we used evolutive harmonic analysis based on the MTM spectral method (Meyers, 2014) (Fig. 4). To extract cycles we performed filtering using the Gaussian filter in the freeware *Analyseries* (Paillard *et al.*, 1996).

The effect of geologic timescales on the estimated periods was tested in two manners. First, we calibrated SL record to different geologic timescales (e.g., GTS2004, GTS2008, GTS2012 and GTS2016). GTS2008 is a slightly modified version of GTS2004, and GTS2016 is a slightly modified version of GTS2012. We then performed spectral analysis on the calibrated SL records to look for the persistence (or not) of the interpreted periods (Supplementary material SM-1). Second, in a more statistical way, we tested the impact of GTS2016 timescale uncertainties on spectral estimates using Monte-Carlo age-random simulations based on Bayesian Markov Chain Monte Carlo (MCMC) simulations (Fig. 2) via 'Bchron' program (Haslett and Parnell, 2008).

We have adapted 'Bchron' program in order to satisfy the required conditions: (1) Continuity and monotony of the age scale, and (2) age variance increases away from anchor age points. Ages and uncertainties of Stage boundaries are from the International Commission on Stratigraphy website.

To test the significance of long-term SL change in terms of climate forcing, we used stable oxygen isotopes ($\delta^{18}O$) data (Supplementary material SM-1), which were correlated to SL using cross-MTM spectral analysis (Huybers and Denton, 2008). We used the cross-MTM spectral analysis to study the coherency and phase relationship between $\delta^{18}O$ and SL signals at the 36 Myr cycle band.

Based on the classical model of Paczynski (1990), we have build a parametrized model for the Milky Way with several free parameters that have then been adjusted in order to fit the observed periodicities of 72 and 254 Myr (Supplementary material SM-2). Additionally, we have modeled the GCR flux using the sources distribution of Lorimer (2004). The numerical integration of the orbit of the Sun then provides the variation of GCR flux on Earth in the past. This target curve closely matches the SL curve.

## 3. Results

*3.1. Time-series analysis of the Phanerozoic sea-level data*

The Phanerozoic eustatic curve shows two strong megacycles (Fig. 1): roughly spanning the Ceno-Mesozoic (0–251 Ma) and the Paleozoic (251–542 Ma), with superimposed shorter prominent oscillations. Thus, the Ceno-Mesozoic megacycle contains 7 shorter oscillations of ~36 Myr mean duration (C1-C2 and M1 to M5 cycles).

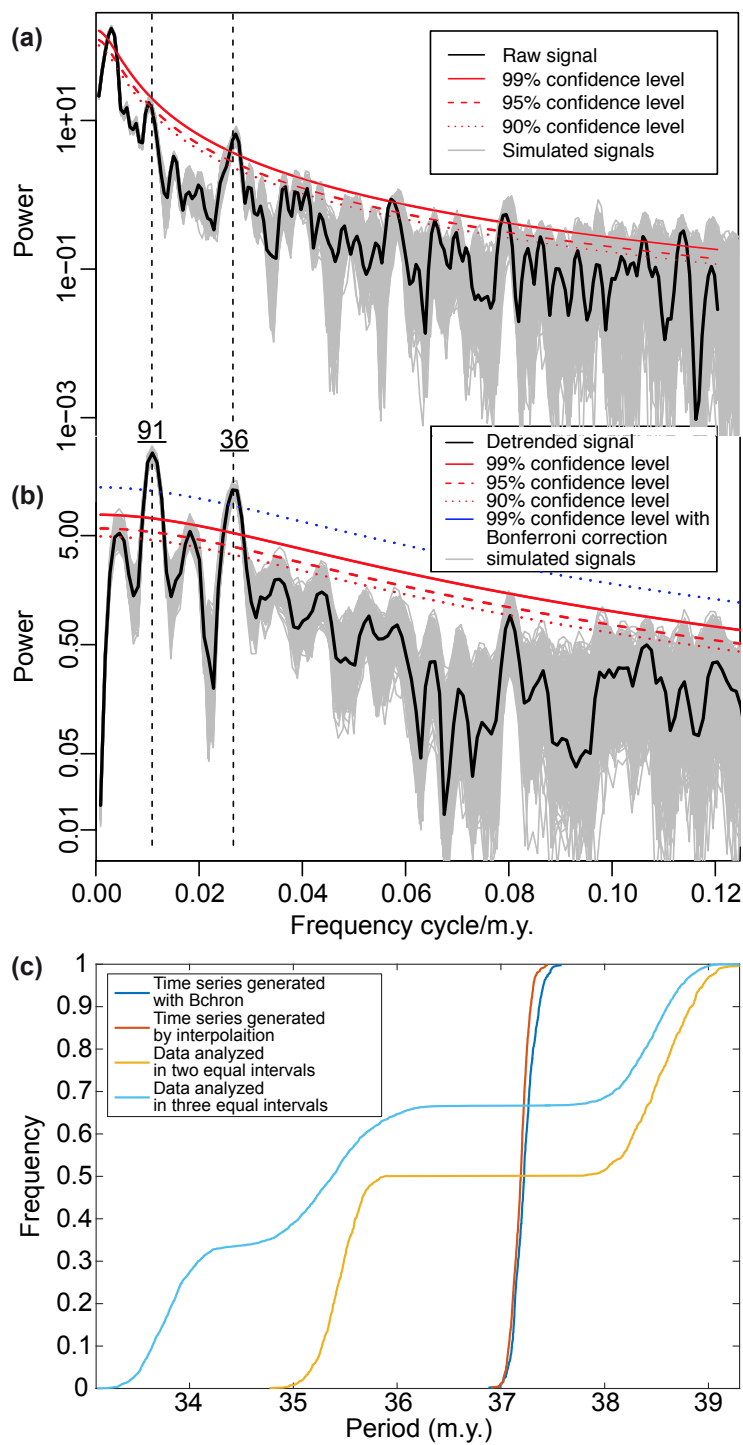

**Fig. 2** Spectral analysis of the Phanerozoic SL data. **(a)** 2π-MTM power spectrum of the raw Phanerozoic SL data (the data were 2x-padded prior spectral analysis, see Supplementary material SM-1 for details of spectral analysis), results of robust red noise modeling were estimated using linear fitting and median filtering over 20% of the Nyquist frequency. Grey-colored spectra are 1,000 Markov Chain Monte Carlo (MCMC) simulations. **(b)** as in 'a' where spectral analysis was applied to the detrended data (6[th]-order polynomial fit removed, see Fig. 1b). We also added Bonferroni correction for the 99% confidence level. **(c)** Experimental cumulative distribution function of the 36 Myr period of SL data over three time intervals: the entire SL data, SL data split into two equal time intervals, SL data split into three equal time intervals.

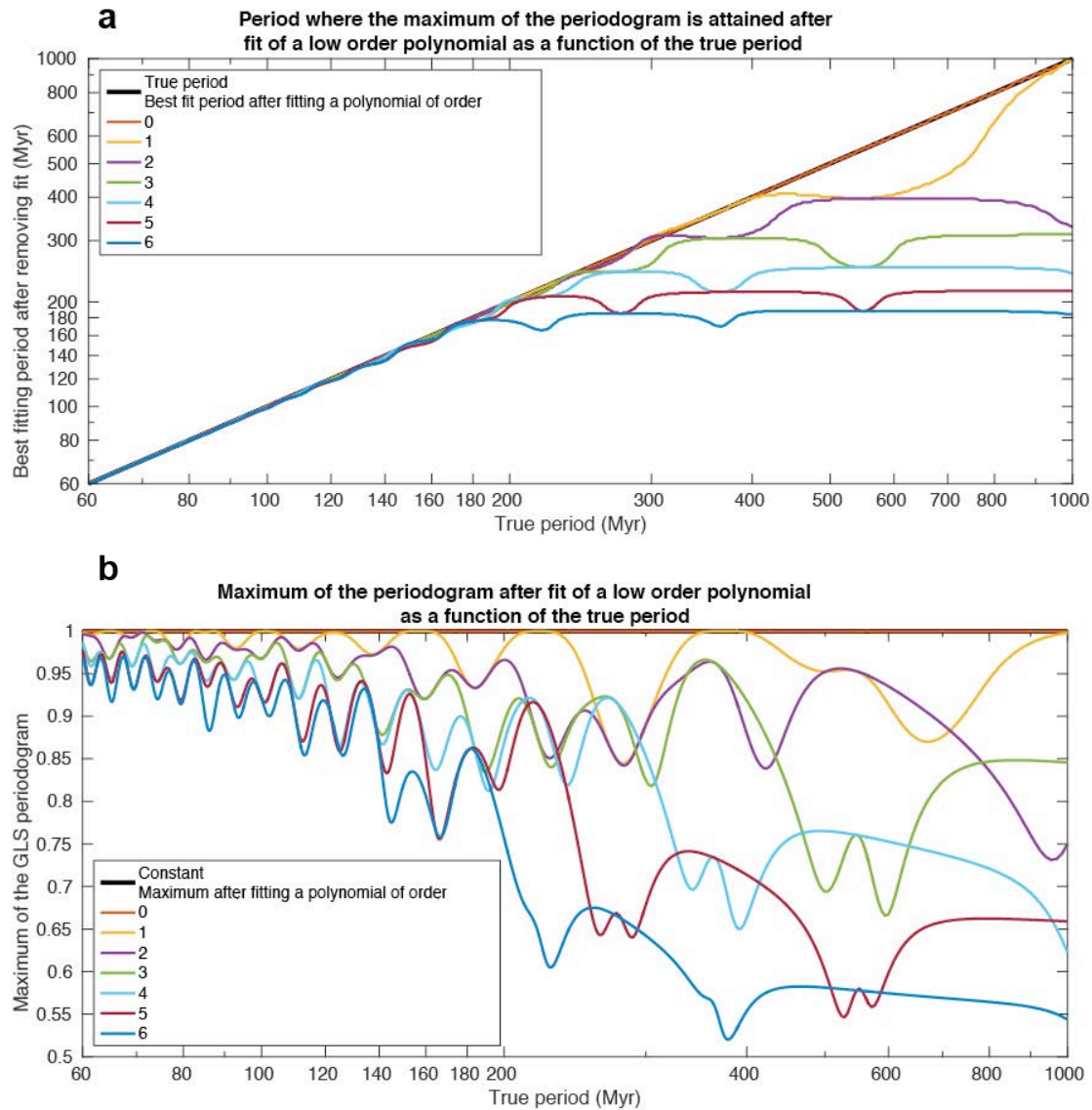

**Fig. 2** The effect of long-term detrend on spectral period estimates. **(a)** Change in spectral period estimates after successive polynomial fits. **(b)** Ratio of amplitude maxima of Generalized Lomb-Scargle (GLS) periodogram before and after polynomial fits. Note that spectral outputs of periods shorter than 180 Myr are not affected by the detrend process.

Spectral analysis of the Phanerozoic SL data shows three significant peaks of ~9.3 ~36 and ~91 Myr periods (Figs. 2, SM-1-1, SM-1-2). The ~9.3 My peak corresponds to the shorter 2$^{nd}$ order eustatic sequences of Haq *et al.* (1987) (i.e., their "*supersequences*"), which were shown to be quasi-periodic for, e.g., the past ~70 Myr. A cyclicity of ~9.3 Myr period was first highlighted in the Cenozoic carbon-isotopic data (Boulila et al., 2012).

The ~36 Myr peak represents a prominent and continuous cyclicity throughout the Phanerozoic. Although this periodicity was not interpreted as continuous in previous studies (Haq *et al.*, 1987), it sometimes matches the longer 2$^{nd}$ order eustatic sequence boundaries (i.e., "*supersequence sets*", i.e., , Tejas-A, Tejas-B, Zuni-A, Zuni-B, etc., Fig. SM-1-3), but at other times shows correspondance to doublets of these (for example Zuni-C and lower Absaroka, Fig. SM-1-3) (Haq *et al.*, 1987). The doublets are likely harmonics of the main ~36 Myr cycle. The ~91 Myr peak corresponds to the average of harmonics (two or three) of the ~36 Myr cycles (SM-1). This period may have a galactic (climatic) and/or a tectonic origin given its important amplitude in the SL record. We, however, explore the relatively well recognized ~36 Myr cyclicity in terms of solar-system vertical period. Evolutive Harmonic Analysis (EHA) shows evidence for the continuity of the ~36 Myr cyclicity throughout the Phanerozoic (Fig. 4). Thus, our time-series analysis of the Phanerozoic eustatic curve shows a strong continuous ~36 Myr cyclicity superimposed on the Ceno-Mesozoic and Paleozoic megacycles. The ~9.3 and ~91 Myr cyclicities are not continuously recorded within the SL signal compared to the exceptionally continuous ~36 Myr cyclicity (Fig. 4). It is likely that the unstationarity of the ~91 Myr cycle could also be related to the length of the time series, which is too short to precisely detect such cyclicity.

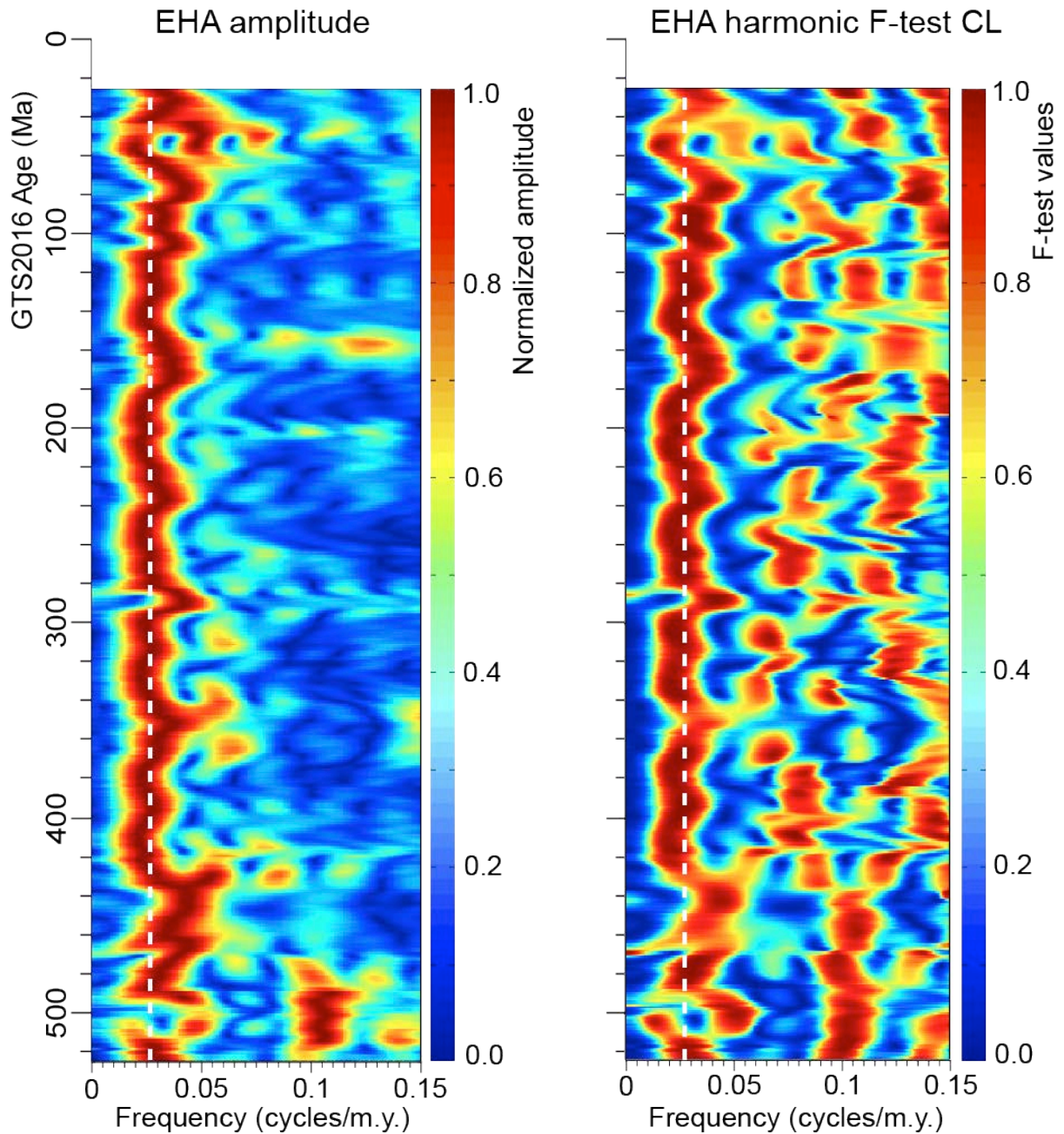

**Fig. 4** Evolutive harmonic analysis (EHA) using 2π-MTM spectral analysis (Meyers, 2014) of the raw Phanerozoic eustatic data. Left panel: amplitude spectrogram (window = 50 Myr, step = 1 Myr). Right panel: the F-test confidence levels. Note the prominent, significant cyclicity at the mean period of ~36 Myr (vertical dashed white line) throughout the Phanerozoic Eon.

*3.2. Astronomical modeling: solar system motions and galactic cosmic rays*

We utilize a quantitative approach: starting from a well-known model of galactic potential (Paczynski, 1990), we have modified the free parameters of the model in order to fit the observed geological periodicities to the orbital motion. We demonstrate that both periods of 72 and 254 Myr can be interpreted as the vertical and radial periods in a galactic model that fulfill the other observational constraints (Fig. 1c,e and SM-1). The ~36 Myr SL cycles are then interpreted to represent the half-period of the 72 Myr vertical oscillations of the solar system moving in and out of the galactic plane.

In addition, we have modeled the distribution of GCR and demonstrated that the variation of incidental GCR on Earth is well correlated with the inferred climate changes from the geological record (Fig. 1).

This correlation is in accord with the notion that increased clouds and planetary albedo from GCR enhancement during times of galactic midplane crossings and galactic-center nearing lead to climatic coolings (Shaviv and Veizer, 2003; Gies and Helsel, 2005), thereby inducing large amplitude glacio-eustatic SL minima. The highest amplitudes of the eustatic megacycles related in the model to the radial motion goes with the idea that extended exposure to the higher GCR flux associated with the nucleus can lead to increased and extensive cloud cover and long ice-age epochs on Earth (Fig. 1a-c).

## 4. Discussion

*4.1. A proposed new hierarchy of long-term SL change and potential drivers*

The seminal studies showing the hierarchy of global sea-level (eustatic) sequences from seismic data and well-log stratigraphy, and their impact on sedimentary deposition were undertaken by researchers at Exxon Production Research Company (Vail et al., 1977; Haq et al., 1987 and 1988). In particular, Vail et al. (1977) divided these depositional sequences temporally into six orders ranging from tens-hundreds of millions years (first- and second-order) to tens of thousands years (sixth order). First- and second-order SL sequences were ascribed to tectono-eustatic changes in the global ocean volume, while fourth-, through sixth-order SL sequences were attributed to climate change within the Milankovitch (insolation) band. However, third-order SL sequences were interpreted as the result of climate or tectonic forcing (Vail et al., 1991; Cloetingh, 1988; Strasser et al., 2000), while more recent studies have argued long-period (1.2 and 2.4 Myr) Milankovitch forcing for the

Cenozoic and Mesozoic third-order SL cycles (Boulila et al., 2011).

In addition, Haq et al. (1987) proposed a detailed subdivision at the first- and second-order sequences as follows. They subdivided the second-order into two suborders, longer (or *megacycle set*) and shorter (*megacycle*), and similarly the first-order into two suborders, longer (*supercycle set*) and shorter (*supercycle set*). Although durations of each suborder could vary considerably (Section 3.1), possibly because such subdivision would require statistical treatment which takes into accound both the amplitude and the period of the signal, we should note that Haq et al. (1987) manually arrived at the same number of (sub)orders detected today by spectral analysis.

Indeed, time-series analysis of long-term Phanerozoic SL data indicates mainly four frequency bands (Section 3.1, Table 1): ~9.3 Myr, ~36 Myr, ~91 Myr and ~250 Myr. Therefore, we suggest that the ~9.3 and ~36 Myr periodicities correspond respectively to shorter and longer second suborders. The ~250 Myr period correspond to the longer first suborder. The ~91 Myr mean period, which is temporally not well constrained in previous studies, could correspond to the shorter first suborder (Table 1).

These very long-term SL variations have been generally attributed to tectonic changes through geodynamically induced changes in the volume of the ocean basins (Vail et al., 1977). The attribution of these SL changes to tectonics rather than climate (or to both) is dictated by the lack of long time series of climatic proxies useful for comparison with eustasy, even though some researchers have pointed out the correspondence between very long-term climatic and eustatic trends (Kaiho and Saito, 1994; Abreu et al., 1998, Haq and Schutter, 2008). Tectonic forcing as the cause for long-term SL change has been generally accepted because major plate tectonic motions occur at a very slow pace, and also because no other explanation has been offered for the high-amplitude eustatic variations, especially during the so-called greenhouse periods when there were extensively no ice sheets. Other researches have postulated that icehouse conditions may have existed throughout the Earth's history and suggest that SL oscillations were largely glacio-eustatically driven (Matthews and Al-Husseini, 2010; Boulila et al., 2011). Unlike the well known climatically-driven higher-frequency SL fluctuations, those related to Milankovitch (insolation) orbital forcing, the mechanisms of low-frequency SL changes still remain controversial (Table 1).

Indeed, very long-term climatic variations (on the time scales of 10s to few 100s of Myr) are thought to be controlled by galactic cosmic ray (GCR) changes via solar-system motions in the galaxy (Shaviv and Veizer, 2003; Gies and Helsel, 2005; Svensmark, 2006). In particular, the prominent ~36 Myr cyclicity that we have continuously detected throughout the Phanerozoic SL record has received increasing attention since 1980's from multiple disciplines: astronomy (e.g., Svensmark, 2006; Bailer-Jones, 2009), cosmoclimatology (e.g., Svensmark, 2007), geology (Kaiho and Saito, 1995), and astrobiology (Raup and Sepkoski, 1984; Medvedev and Melott, 2007).

Equivalent periodicity has been predicted in the motion of the solar system when it moves down- and upwards the galactic midplane. This would induce significant change in GCR and the resulting climate change (e.g., Svensmark, 2006). In addition, the ~250 Myr megacycles seem to modulate the ~36 Myr cyclicity. In this context, we are tempted to interpret the ~36 and ~250 SL cyclicities as astronomical in origin (Section 3.2) and investigate their possible climatic significance (Section 4.1), but without excluding the obvious potential tectonic drivers of climatic and SL changes (Section 4.2). Finally, a possible coupling between tectonics and astronomy on Earth's climate change as feedbacks is quite likely (e.g., Huybers and Langmuir, 2009; Kutterolf et al., 2012; Crowley et al., 2015).

| Order | Suborder | Mean period (Myr) | Causal mechanism | Astronomy |
|---|---|---|---|---|
| First | Longer | 250-300* | Tectonic, galactic | Radial motion? |
| First | Shorter | 91* | Tectonic, galactic? | |
| Second | Longer | 36* | Tectonic, galactic | Vertical motion |
| Second | Shorter | 9.3 | Milankovitch? | Eccentricity? |
| Third | Longer | 2.4 | Milankovitch | Eccentricity |
| Third | Shorter | 1.2 | Milankovitch | Obliquity |
| Fourth | | 0.405 | Milankovitch | Eccentricity |
| Fifth | Longer | 0.17 | Milankovitch | Obliquity |
| Fifth | Shorter | 0.10 | Milankovitch | Eccentricity |
| Sixth | Longer | 0.04 | Milankovitch | Obliquity |
| Sixth | Shorter | 0.02 | Milankovitch | Precession |

**Table 1:** Sea-level (SL) hierarchical orders inferred from time-series analysis of the Phanerozoic SL record. Note that Milankovitch and Galactic forcings act on SL via climatically (insolation and galactic cosmic rays respectively) driven glacio-eustasy and/or thermo-eustasy, while tectonic forcing acts on the changes in ocean basin volume. The proposed third- to sixth orders are from Boulila et al. (2011), while the first and second orders are an update from the present study (see Section 4.1).
The 0.17 Myr Milankovitch (obliquity modulation) cyclicity is recently recorded over nearly ten million years in the middle Eocene (Boulila et al., 2018).
* Phanerozoic mean periodicity.

### 4.2. Climate significance of the 36 and 250 Myr SL periodicities

Ice-volume proxies provide rough estimates of ice sheets formation through the Phanerozoic (Fig. 1a). Although a good correspondence can be observed between the eustatic minima of the megacycles and larger glacial episodes, no correlation is discernable at the scale of ~36 Myr cycles. Nevertheless, a deep-sea composite $\delta^{18}O$ curve over the past 115 Ma (Zachos *et al.*, 2001, 2008; Cramer *et al.*, 2009; Friedrich *et al.*, 2012) extended to ~200 Ma (Veizer and Prokoph, 2015) shows evidence of a ~33 Myr cycle correlated to its equivalent in the SL record (Figs. 5, SM-1-6, SM-1-7). A ~31 Myr $\delta^{18}O$ cycle over the past 200 Ma was first demonstrated by Svensmark (2006) using a previous $\delta^{18}O$ compilation (Veizer *et al.*, 1999). The $\delta^{18}O$ record over a large part of the Cenozoic indicates both temperature and ice-sheet variations (Zachos *et al.*, 2001). Thus, the ~33 Myr $\delta^{18}O$ cycle may reflect a glacio-eustatically driven SL cyclicity. More recently, Shaviv *et al.* (2014) postulated the presence of the 32 Myr $\delta^{18}O$ cyclicity throughout the Phanerozoic, although data scatter and gaps exist in their original, master ML200, $\delta^{18}O$ signal for the past 490 Ma (Fig. SM-1-5a). In particular, the temporally most constrained 0-65 Ma interval does not show an obvious 32 Myr cyclicity either in their ML200 $\delta^{18}O$ record or in their ML175 record (Fig. SM-1-5b), in contrast to the deep-sea record (Zachos *et al.*, 2001, 2008) that shows evidence of the ~33 Myr cyclicity (Fig. SM-1-6). This again may be an impediment of the data resolution for the very long geological signals, unlike the constrained SL record used here.

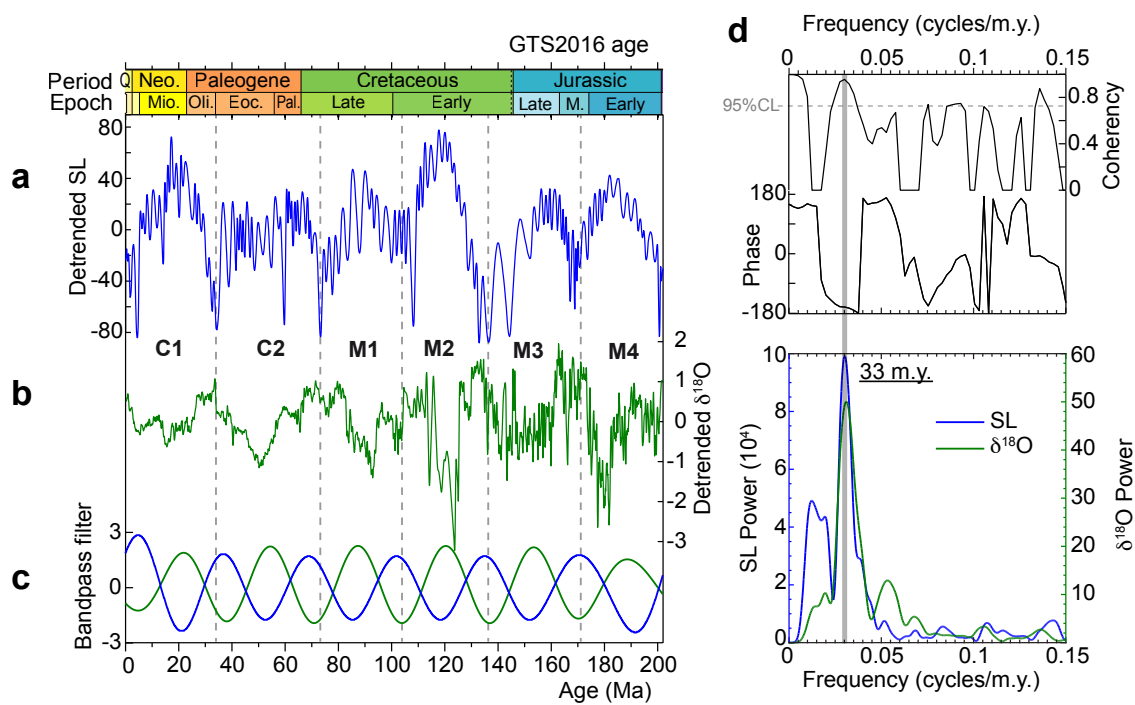

**Fig. 5** Correlation of long-period geological cycles over the past 202 Ma (the mean period is 33 Myr over this time interval, see Fig. 2c). **(a)** Detrended SL (6th-order polynomial fit removed, see Fig. 1b for the raw SL). **(b)** Smoothed (least-square) and detrended stable oxygen isotopes, $\delta^{18}O$ (6th-order polynomial fit removed, see SM-1 for the raw $\delta^{18}O$ data and compilation). **(c)** Bandpass filtered SL (blue) and $\delta^{18}O$ (green) (0.03 ±0.01 cycles/Myr). **(d)** $2\pi$ cross-MTM spectral analysis of detrended SL and $\delta^{18}O$ (both data were 1x-padded prior spectral analysis). Note that both signals are highly coherent (0.96) at the 33 Myr cycle band and nearly anti-phased (-170°), which is the expected phase relationship: heavier $\delta^{18}O$ values (cooler climate) correlate to lower sea levels. Except for 'M2' $\delta^{18}O$ equivalent cycle, which exhibits lower $\delta^{18}O$ data (Fig. SM-1-7) the other 33 Myr $\delta^{18}O$ related cycles are defined with higher resolution.

*4.2. Origin of the 36 and 250 Myr SL periodicities: astronomy or tectonics?*

The similarities of the multi-Myr cyclicities in the geological record and the astronomical model indicates that vertical and radial solar-system motions could induce long-term climatic variations through the modulation (waxing and waning) of ice sheets and the consequent eustatic fluctuations. In particular, vertical and radial motions would modify the proximity of the solar system to density sources, and thus could have induced significant changes in the incidental GCR flux. Enhanced GCR would increase clouds and planetary albedo, climatic cooling (Shaviv and Veizer, 2003), ice age epochs (Gies and Helsel, 2005), and glacio-eustatic SL falls.

Nevertheless, the potential tectonic drivers for the ~36 and ~250 Myr cannot be ignored (Vail *et al.*, 1977; Haq *et al.*, 1987; Kaiho and Saito, 1995; Cloetingh and Haq, 2015), which leads us to reassess the importance of a coupled climate and tectonics effect (Lagabrielle *et al.*, 2009) that could play a role in very long-term SL changes. It is very likely that several drivers (e.g., solar-system cycles, Milankovitch forcing, non-astronomical climatic processes, tectonics), and their combined effects control the eustatic variations at the multi-Myr time scales.

Currently, problems remain in accurate paleo-tectonic reconstructions that hinder the unambiguous determination of the duration and nature of tectonic variations that can produce global effects, except at the large scale of megacycles where good correlations have been inferred between eustasy and well-constrained tectonic processes, e.g., mean age of the oceanic crust and mid-ocean ridge volumes (Cogné *et al.*, 2006).

Our most intriguing result is the detection of significant ~36 Myr cyclicity throughout the Phanerozoic, superimposed on the two ~250 Myr megacycles. These periodicities have been extracted from the most continuous and longest geological (SL) record, compared to those previously used for the detection of similar cyclicities ascribed to the Solar System motions. Moreover, the ~36 Myr cyclicity also emerges in the most constrained Cenozoic time series with a high-resolution ice-volume and temperature proxy data (Fig. 5). Furthermore, the ~250 Myr megacycle minima match the modeled and geologically documented major icehouse windows of the Phanerozoic (Fig. 1). The above close links provide compelling evidence for climatically driven SL at the longer timescales of ~36 and ~250 Myr. Interestingly, the modeled astronomically driven GCR flux is consistent with long-term cooling towards present day, as inferred from multiple paleoclimatic proxies (e.g., Lisiecki and Raymo, 2005). Our present-day proximity to the galactic mid-plane would favor great exposure of the Earth to cosmic rays, inducing a long-term cooling and lower SL. In short, while the ~36 Myr cyclicity could be astro-climatically driven, the megacycles could be astro-climatically and/or geodynamically driven.

Although we draw attention to this potential linkage between astronomy and climate, we must caution that several aspects in this correlation remain conjectural but at the same time present challenging questions that should be investigated in the future. One is the improvement of knowledge of the GCR impact on Earth's climate at these longer timescales (Kirby *et al.*, 2011). The GAIA astrometric mission (Binney, 2002) will soon release new data on galactic parameters that could reduce uncertainties of solar-system periodicities. The resulting improvement of our knowledge of the galactic structure and the vicinity of the Sun in the Milky Way could constrain the ~36 Myr vertical periodicity, which seems robust compared to previous estimates, and condones (or rejects) the ~254 Myr radial periodicity, which is different from previous estimates, but fits the present available galactic parameters.

In this vain, we would like to point out that if the potential correlation presented here between geological and astronomical periodicities would be confirmed, this could provide an exceptional, independent constraint on the history and structure of the Milky Way by providing a record of the past evolution of the orbit of the Sun within the Galaxy over the past 542 Ma.

Finally, the icehouse phases observed throughout the Phanerozoic that more of less match the 250 Myr megacycle (see Fig. 1a) and also the changes in the ocean crustal production rates (Fig. SM1-4) favor tectonically-driven mechanisms through megacyclicity (i.e., the Wilson cycles) embodied in the coalescence-breakup of supercontinents (e.g., Zaffos et al., 2017). Similarly, the observed 36 Myr periodicity in the $d^{18}O$ record could also be plausibly due to tectonically-driven opening and closing of oceanic gateways and changes in circulation patterns that in turn modulate climates (Zachos *et al.*, 2001; Lagabrielle *et al.*, 2009).

## 5. Conclusions

We used a relatively well-constrained Phanerozoic (0–542 Ma) sea level (SL) record to statistically demonstrate a series of frequency bands providing constraints on the eustatic SL hierarchy. The first-order SL cycles are on the order of ~250 Myr although the analysed 542 Myr interval of the past is not enough long to precisely assess its mean periodicity. The second-order SL cycles could be subdivided into two suborders: a longer suborder of ~36 Myr period and a shorter suborder of ~9.3 Myr. A potential periodicity of ~91 Myr does not correspond to any previously defined order, thus we infer it as a shorter suborder of the first-order sequences, and the ~250 Myr being a longer suborder of the same order. While the ~9.3 Myr pseudo-periodicity was previously attributed to long-period Milankovitch modulation cycle, the other orders were generally attributed to major plate tectonic motions, generating changes of ocean volume.

Of particular interest is a prominent and continuous ~36 Myr cyclicity superimposed on the two ~250 Myr megacycles, which are of the same order as those predicted by the Solar System's motions within the Milky Way. The ~36 Myr SL cyclicity has also been detected in the climate proxy of the $\delta^{18}O$ data. Thus, we are tempted to propose a potential connection between climate and SL changes via galactic cosmic rays (GCR) flux for this periodicity. We have constructed a coherent model for the galactic potential and GCR flux that can be correlated to the geological record. At the same time, we do not rule out a possible coupled climate-tectonics effect that could play a role in very long-term SL changes. Currently, problems remain in accurate paleo-tectonic reconstructions that hinder the unambiguous determination of the duration and nature of tectonic periodicities useful for comparison with long-term SL cycles.

**Acknowledgments**

This work was supported by French ANR Grant ASTS-CM, and INSU-SYSTER Grant. S.B. thanks very much O. Friedrich (Heidelberg University), who kindly provided the compiled deep-sea stable isotope data. This study benefitted from discussions and exchanges with J. Binney, W. Dehnen, T. Delahaye, T. Montmerle, B. Famaey K. Kotera, and M. Haywood. We also thank the two anonymous reviewers, who significantly improved the quality of the manuscript. The paper also benefitted from reviews of three experts in time-series analysis (Graham Weedon and two anonymous reviewers) and three experts in astronomy (Henrik Svensmark and two anonymous reviewers).
**References**

Abreu, V.S, Hardenbol, J, Haddad, G.A, Baum, G.R,, Droxler, A.W, Vail, P.R., 1998. Oxygen isotope synthesis: A Cretaceous ice-house?  Mezozoic and Cenozoic Sequence Stratigraphy of European Basins, eds de Graciansky PC, Jacquin T, Hardenbol, SEPM Special Publication (Soc Sediment Geol, Tulsa, OK) 60:75–80.

Bahcall, J.N., Bahcall, S., 1985. The Sun's motion perpendicular to the galactic plane. Nature, 316, 706–708.

Bailer-Jones, C.A.L., 2009. The evidence for and against astronomical impacts on climate change and mass extinctions: A review. Int. J. Astrobiol., 8, 213–239.

Binney, J., 2002. Components of the Milky Way and GAIA. EAS Publications Series, 2, 245–256, doi:10.1051/eas:2002023.

Boulila, S., Galbrun, B., Laskar, J., Pälike, H., 2012. A ~9 myr cycle in Cenozoic $\delta^{13}C$ record and long-term orbital eccentricity modulation: Is there a link? Earth Planet. Sci. Lett. 317–318, 273–281.

Boulila, S., Galbrun, B., Miller, K.G., Pekar, S.F., Browning, J.V., Laskar, J., Wright, J.D., 2011a. On the origin of Cenozoic and Mesozoic "third-order" eustatic sequences. Earth Science Reviews 109, 94−112.

Boulila, S., Vahlenkamp, M., De Vleeschouwer, D., Laskar, J., Yamamoto, Y., Pälike, H., Kirtland Turner, S., Sexton, P.F., Cameron, A., 2018. Towards a robust and consistent middle Eocene astronomical timescale. Earth Planet. Sci. Lett. 486, 94–107.

Cloetingh, S., 1988. Intraplate stresses: a tectonic cause for third-order cycles in apparent sea level? Society of Economic Paleontologists and Mineralogists Special Publication 42, 19–29.

Cloetingh, S., Haq, B.U., 2015. Inherited landscapes and sea level change. Science, 347, doi: 10.1126/science.1258375 (2015).

Cogné, J.P., Humler, E., Courtillot, V., 2006. Mean age of oceanic lithosphere drives eustatic sea-level change since Pangea breakup. Earth Planet. Sci. Lett., 245, 115–122.

Cramer, B.S., Toggweiler, J.R., Wright, J.D., Katz, M.E., Miller, K.G., 2009. Ocean overturning since the Late Cretaceous: Inferences from a new benthic foraminiferal isotope compilation. Paleoceanography, 24, PA4216, doi: 10.1029/2008PA001683.

Crowley, J.W., Katz, R.F., Huybers, P., Langmuir, C.H., Park, S-H., 2015. Glacial cycles drive variations in the production of oceanic crust. Science 347, 1237–1240.

DeCelles, P.G., Ducea, M.N., Kapp, P., Zandt, G., 2009. Cyclicity in Cordilleran orogenic systems. Nature Geoscience 2, 251–257.

Frakes, L.A., Francis, J.E., Syktus, J.I., eds., 1992. Climate Modes of the Phanerozoic: The History of the Earth's Climate Over the Past 600 Million Years. Cambridge University Press, Cambridge.

Friedrich, O., Norris, R.D., Erbacher, J., 2012. Evolution of middle to Late Cretaceous oceans—A 55 m.y. record of Earth's temperature and carbon cycle. Geology, 40, 107–110.

Ghil, M., Allen, R.M., Dettinger, M.D., Ide, K., Kondrashov, D., Mann, M.E., Robertson, A., Saunders, A., Tian, Y., Varadi, F., Yiou, P., 2002. Advanced spectral methods for climatic time series. Review of Geophysics, 40 (1), 3.1–3.41.

Gies, D.R., Helsel, J.W., 2005. Ice Age Epochs and the Sun's Path through the Galaxy. The Astrophysical Journal, 626, 844–847.

Haq, B.U., Al-Qahtani, A-M., 2005. Phanerozoic cycles of sea-level change on the Arabian Platform. GeoArabia, 10, 127–160.

Haq, B.U., Hardenbol, J., Vail, P.R., 1987. Chronology of fluctuating sea levels since the Triassic. Science, 235, 1156–1167.

Haq, B.U., Hardenbol, J., Vail, P.R., 1988. Mesozoic and Cenozoic chronostratigraphy and cycles of sea-level change. Society of Economic Paleontologists and Mineralogists, v. 42, p. 71-108.

Haq, B.U., Schutter, S.R., 2008. A chronology of Phanerozoic Sea-Level Changes. Science, 322, 64–68.

Haslett, J., Parnell, A.C., 2008. A simple monotone process with application to radiocarbon-dated depth chronologies. Appl. Statist., 57, Part 4, 399–418.

Hilgen, F.J., 2010. Astronomical dating in the 19th century. Earth-Science Reviews, 98, 65–80.

Huybers, P., Denton, G., 2008. Antarctic temperature at orbital time scales controlled by local summer duration. Nature Geoscience, 1, 787–792.

Huybers, P., Langmuir, C., 2009. Feedback between deglaciation, volcanism and atmospheric $CO_2$. Earth Planet. Sci. Lett. 286, 479–491.

Kaiho, K., Saito, S., 1994. Oceanic crust production and climate during the last 100 Ma. Terra Nova, 6, 376–384.

Kirby et al., 2011. Role of sulphuric acid, ammonia and galactic cosmic rays in atmospheric aerosol nucleation. Nature, 476, 7361, 429–433.

Kutterolf, S., Jegen, M., Mitrovica, J.X., Kwasnitschka, T., Freundt, A., Huybers, P. A detection of Milankovitch frequencies in global volcanic activity. Geology 41, 227–230.

Lagabrielle, Y., Goddéris, Y., Donnadieu, Y., Malavieille, J., Suarez, M., 2009. The tectonic history of Drake Passage and its possible impacts on global climate. Earth Planet. Sci. Lett., 279, 197–211.

Laskar J., Fienga A., Gastineau M., Manche H., 2011. La2010: A new orbital solution for the long term motion of the Earth. Astron. Astrophys. 532, A89, doi: 10.1051/0004-6361/201116836.

Laskar J., Robutel P., Joutel F., Gastineau M., Correia A.C.M., Levrard B., 2004. A long-term numerical solution for the insolution quantities of the Earth. Astronomy and Astrophysics, 428, 261–285.

Lisiecki, L.E., Raymo, M.E., 2005. A Pliocene-Pleistocene stack of 57 globally distributed benthic $\delta^{18}O$ records. Paleoceanography, 20, PA1003, doi:10.1029/2004PA001071.

Lorimer, D.R., 2004. The galactic population and birth rate of radio pulsars, in Young Neutron Stars and their Environments. In: Camilo, F., Gaensler, B., Eds., Young neutron stars and their environments. IAU Symposium, 218, 105-112.

Mattews, R.K., Al-Husseini, M.I., 2010. Orbital-forcing glacio-eustasy: A sequence stratigraphic time scale. GeoArabia 15, 129−142.

Medvedev, M.V., Melott, A.L., 2007. Do Extragalactic Cosmic Rays Induce Cycles in Fossil Diversity? The Astrophysical Journal, 664, 879–889.

Meyers, S.R., 2014. Astrochon: An R package for astrochronology. Available at cran.rproject. org/web/packages/astrochron/index.html.

Meyers, S.R., Peters, S.E., 2011. A 56 million year rhythm in North American sedimentation during the Phanerozoic. Earth Planet. Sci. Lett., 303, 174–180.

Milankovitch, M., 1941. Kanon der Erdbestrahlung und seine Anwendung auf das Eiszeitenproblem. Royal Serbian Academy, Section of Mathematical and Natural Sciences, Belgrade, p. 633 [and 1998 reissue in English: Canon of Insolation and the Ice-Age Problem. Belgrade: Serbian Academy of Sciences and Arts, Section of Mathematical and Natural Sciences, 634 pp.].

Ogg, J.G., Ogg, G., Gradstein, F.M., 2016. A Concise Geologic Time Scale 2016. Elsevier, Amsterdam.

Paczynski, B., 1990. A test of the galactic origin of gamma-ray bursts. The Astrophysical Journal, 348, 485–494.

Paillard, D., Labeyrie, L., Yiou, P., 1996. Macintosch program performs timeseries analysis. Eos, 77, 379.

Pälike, H., Norris, R.D., Herrle, J.O., Wilson, P.A., Coxall, H.K., Lear, C.H., Shackleton, N.J., Tripati, A.K., Wade, B.S., 2006a. The Heartbeat of the Oligocene Climate System. Science 314, 1894−1898.

Randal, L., Reece, M., 2014. Dark Matter as a Trigger for Periodic Comet Impacts. Physical Rev. Lett., 112, 161301.1–161301.5.

Raup, D.M., Sepkoski, J.J.Jr., 1984. Periodicity of extinctions in the geologic past. Proc. Nat. Acad. Sci. USA, 81, 801-805.

Ridgwell, A.A, 2005. Mid-Mesozoic Revolution in the regulation of ocean chemistry. Marine Geology, 217, 339–357.

Rohde, R.A., Muller, R.A., 2005. Cycles in fossil diversity. Nature, 434, 208–210.

Schönrich, R., Binney, J., Dehnen, W., 2010. Local kinematics and the local standard of rest. Monthly Notices Roy. Astr. Soc., 403 (4), 1829–1833.

Shaviv, N., 2002. Cosmic ray diffusion from the galactic spiral arms, iron meteorites, and a possible climatic connection. Physical Rev. Lett., 89, 051102.1-051102.4.

Shaviv, N.J., Prokoph, A., Veizer, J., 2014. Is the Solar System's Galactic Motion Imprinted in the Phanerozoic Climate? Scientific Reports, 4, 6150, doi:10.1038/srep06150.

Shaviv, N.J., Veizer, J., 2003. Celestial driver of Phanerozoic climate? GSA Today, 13, 4–10.

Stothers, R.B., 1998. Galactic disk dark matter, terrestrial impact cratering and the law of large numbers. Monthly Notices Roy. Astr. Soc., 300, 1098–1104.

Strasser, A., Hillgärtner, H., Hug, W., Pittet, B., 2000. Third-order depositional sequences reflecting Milankovitch cyclicity. Terra Nova 12, 303−311.

Svensmark, H., 2006. Imprint of Galactic dynamics on Earth's climate. Astron. Nachr./AN, v. 327, 9, 866–870.

Svensmark, H., 2007. Cosmoclimatology: a new theory emerges. Astronomy & Geophysics, 48, 1.18–1.24.

Svensmark, H., 2012. Evidence of nearby supernovae affecting life on Earth. Monthly Notices Roy. Astr. Soc., 423 (2), 1234-1253.

Vail, P.R., Mitchum, R.M., Todd, J.R.G., Widmier, J.M., Thompson, S., Sangree, J.B., Bubb, J.N., Hatlelid, W.G., 1977. Seismic stratigraphy and global changes of sea level. In: C.E. Payton, C.E. Ed., Seismic Stratigraphy – Applications to Hydrocarbon Exploration. Mem. AAPG, 26, 49–212.

van Dam, J.A., Abdul, Aziz, H., Sierra, M.A.A., Hilgen, F.J., van den Hoek Ostende, L.W., Lourens, L.J., Mein, P., van der Meulen, A.J., and Pelaez-Campomanes, P., 2006. Long-period astronomical forcing of mammal turnover. Nature 443, 687–691.

Veizer, J., Ala, D., Azmy, K., Bruckschen, P., Buhl, D., Bruhn, F., Carden, G.A.F., Diener, A., Ebneth, S., Godderis, Y., Jasper, T., Korte, C., Pawellek, F., Podlaha, O., Strauss, H., 1999. $^{87}Sr/^{86}Sr$, $\delta^{13}C$ and $\delta^{18}O$ evolution of Phanerozoic seawater. Chem. Geol., 161, 59–88.

Veizer, J., Prokoph, A., 2015. Temperatures and oxygen isotopic composition of Phanerozoic oceans. Earth-Science Reviews, 146, 92–104.

Zachos, J.C., Dickens, G.R., Zeebe R.E., 2008. An early Cenozoic perspective on greenhouse warming and carbon-cycle dynamics. Nature, 451, 279–283.

Zachos, J.C., Pagani, M., Sloan, L., Thomas, E., Billups, K., 2001. Trends, Rhythms, Aberrations in Global Climate 65 Ma to Present. Science, 292, 686–693.

Zaffos, A., Finnegan, S., Peters, S.E., 2017. Plate tectonic regulation of global marine animal diversity. Proc. Nat. Acad. Sci. 114, no. 22, 5653–5658.

Zechmeister, M., Kürster, M., 2009. The generalised Lomb-Scargle periodogram. A new formalism for the floating-mean and Keplerian periodograms. Astronomy & Astrophysics 496, 577–584.

# SUPPLEMENTARY MATERIAL (SM1 & SM2)

# Long-Term Cyclicities in Phanerozoic Sea-Level Sedimentary Record and Their Potential Drivers

**Slah Boulila[a,b,1], Jacques Laskar[b], Bilal U. Haq[a,c], Bruno Galbrun[a], and Nathan Hara[b]**

[a]Sorbonne Universités, UPMC Univ Paris 06, CNRS, Institut des Sciences de la Terre de Paris (ISTeP), 4 place Jussieu 75005 Paris, France; [b]Astronomie et Systèmes Dynamiques IMCCE, Observatoire de Paris, 77 Avenue Denfert-Rochereau, 75014 Paris, France; and [c]Smithsonian Institution, Washington DC, USA.

[1]Corresponding author. Tel: +33.144274163; Fax: +33.144273831. E-mail: slah.boulila@upmc.fr.

## *Supplementary material SM1*

**I. Supporting information on time-series analysis**

Spectral analysis of both undetrended and detrended Phanerozoic eustatic data (Fig. S1a,b) shows two significant (>99% CL) peaks centered on ~36 and ~91 Myr A less significant (>95% CL) peak centered on ~9.3 Myr is also present. While the ~36 Myr cyclicity can be visually examined, neither the ~9.3 Myr nor the ~91 Myr are visually obvious. The ~9.3 Myr persists even when considering different timescales (Fig. S2). It may correspond to the shorter 2nd order eustatic sequences of Haq et al. (ref. 1) (i.e., their *"supersequences"*, Fig. S3), which were shown to be quasi-periodic e.g., in the past ~70 Ma. A cyclicity of ~9 Myr period was highlighted, for the first time, in the Cenozoic carbon-cycle proxies (ref. 2).

To further pursue the ~91 Myr peak we performed spectral analysis per intervals and applied filtering. Spectra of both 0-251 Ma (Cenozoic-Mesozoic) and 251-542 Ma (Paleozoic) intervals (Fig. S1c,d) reveal the persistence of the ~36 Myr peak, however, the ~91 Myr is 98 Myr in the Ceno-Mesozoic and only 86 Myr in the Paleozoic. Bandpass filter output of the ~36 Myr cycle band points to the persistence of this cycle at least throughout the last 500 Ma (Fig. S3). This result is strongly supported by evolutive harmonic analysis (main Fig. 4). For the ~91 Myr 'unstable' peak, we applied filtering in two different ways, first to the whole (0-542 Ma) data, then, per intervals: the Ceno-Mesozoic (0-251 Ma) and the Paleozoic (251-542 Ma) intervals. Filtering the whole data is more conservative than that per intervals (filter-edge effects). Thus, we provide filter output of the whole data (Fig. S3) to interpret the two aforementioned peaks (~98 and ~86 Myr, Fig. DR1c,d). Ceno-Mesozoic filter output indicates that the ~98 Myr peak corresponds to the average of the two following bundlings: the three (C2-M1-M2) and the three (M3-M4-M5) ~36 Myr cycles (Fig. DR2). Paleozoic filter output points to the same conclusion (i.e., 36 Myr harmonics). The ~86 Myr peak equals to the average of the three following bundlings : the two (P1-P2), the two (P3-P4), and the three (P5-P6-P7) ~36 Myr cycles (Fig. S3). In summary, filtering shows that the ~91 Myr peak may originate from harmonics of the fundamental ~36 Myr cycle, as the average of two or three cycles. The unstationary of the ~91 Myr cycle could also be related to the length of the time series, which is too short to precisely detect such cyclicity. Despite the possibility that the ~91 Myr cycle could have a galactic (and thus climatic) significance, we will focus on the relatively well constrained ~36 Myr (see main text).

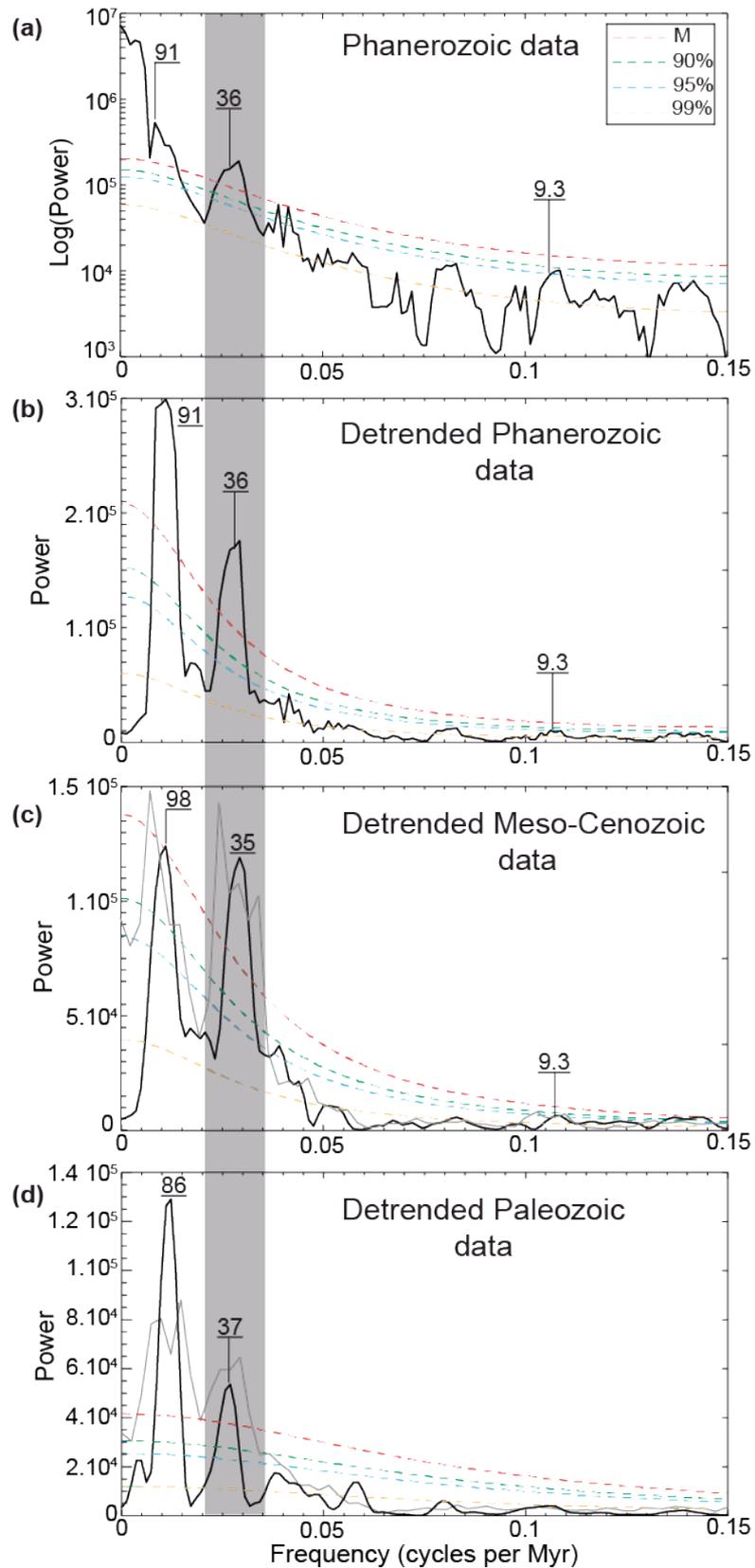

**Figure S1:** 2pi-MTM power spectra of the Phanerozoic eustatic (ref. 1,3,4). Results of noise modeling were estimated using linear fitting and median filtering over 20% of the Nyquist frequency. **(a)** The original whole (Phanerozoic) data (Fig. 1b in the main text). **(b)** The detrended whole data (Fig. 1d in the main text). **(c)** The detrended Cenozoic-Mesozoic (0-251 Ma) interval without (gray spectrum) and with (black spectrum) 1x-zero padding of the series. **(d)** The detrended Paleozoic (251-542 Ma) interval without (gray spectrum) and with (black spectrum) 1x-zero padding of the series. Zero-padding is used to precisely determine periods of spectral peaks. The ~36 Myr cyclicity depicted by the continuous spectral peak throughout the Phanerozoic Eon is shown by gray shaded area (see also evolutive harmonic analysis in main Fig. 4 for the continuity of the ~36 Myr cyclicity).

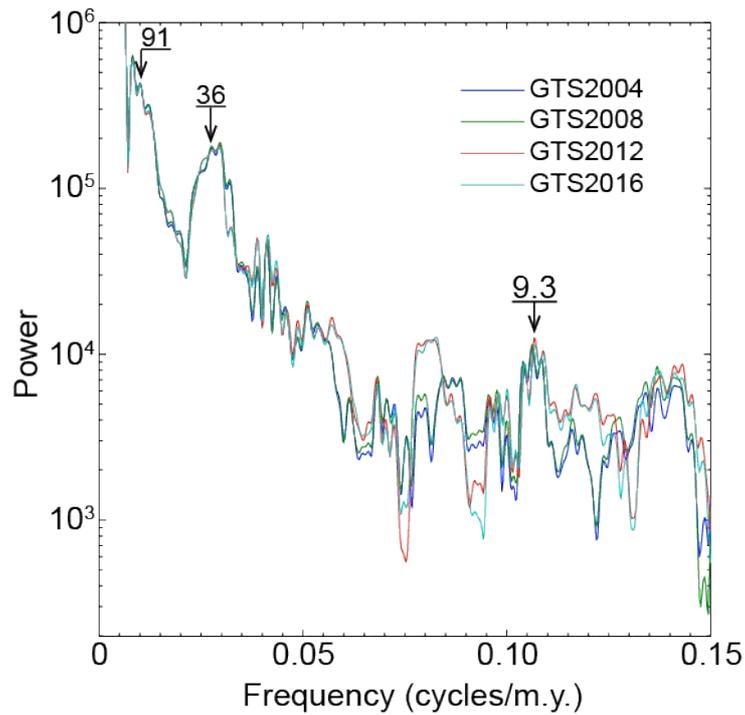

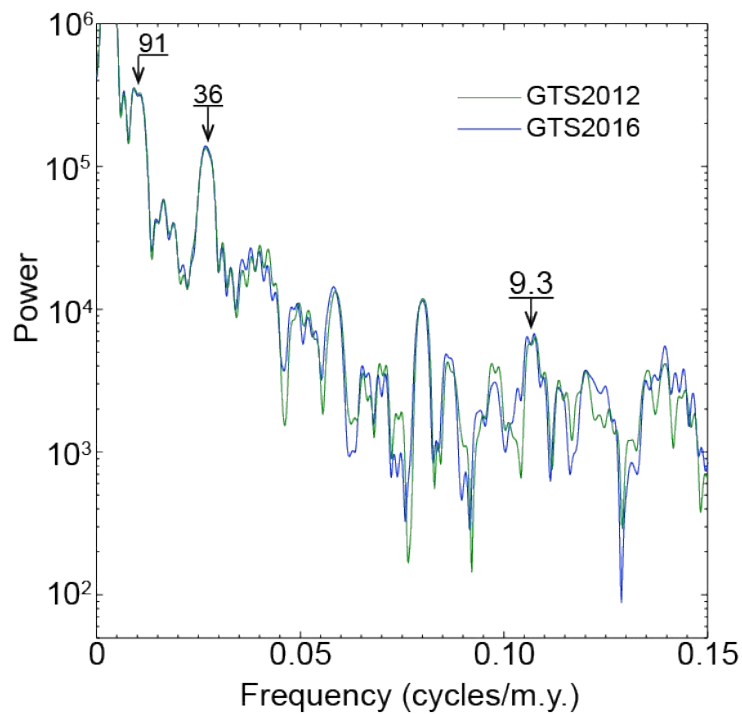

**Figure S2:** 2 π-MTM power spectra of the raw Phanerozoic eustatic data using different Geologic timescales GTS. Upper spectra: spectra without padding. Lower spectra: spectra of the 2x padded data. Note the persistence of the lower frequencies whatever the GTS version. We have also tested GTS2016 uncertainties on the lower frequencies using Markov Chain Monte Carlo (MCMC) Bchron simulations (see main Fig. 2).

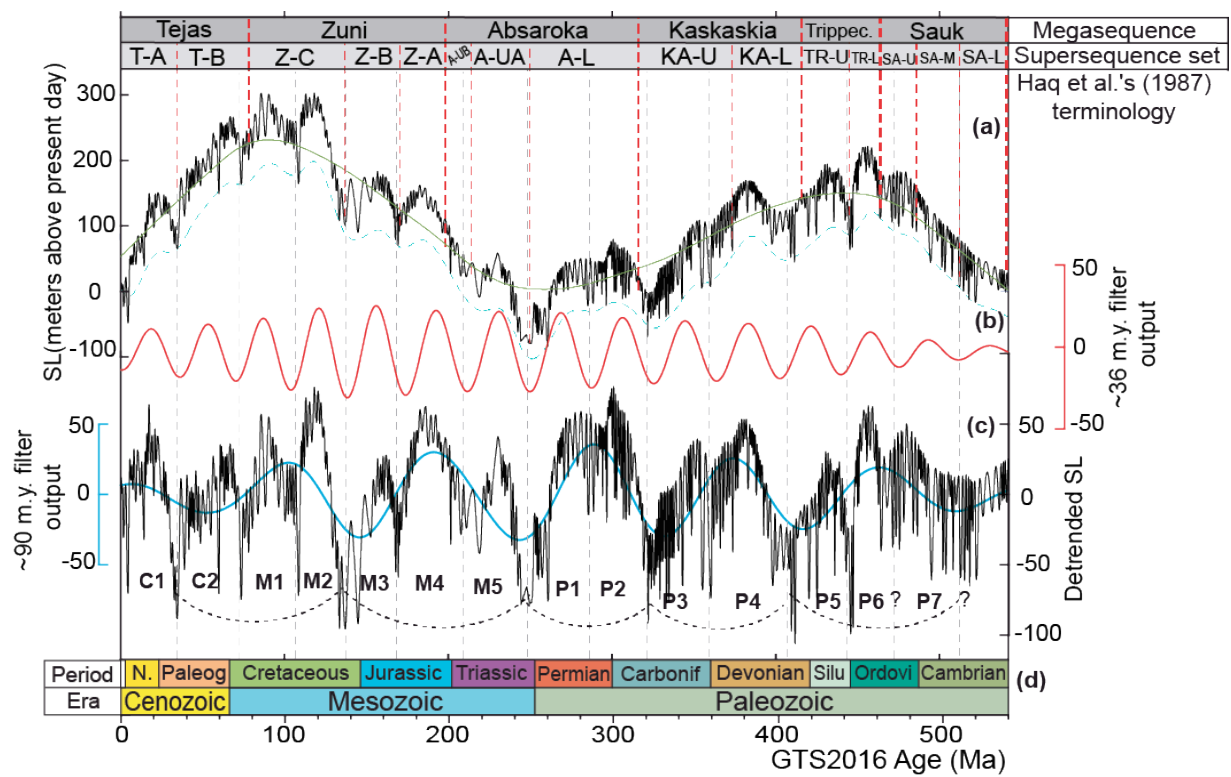

**Figure S3: (a)** Phanerozoic eustatic curve (ref. 1,3,4) with the fitted megacycles and Haq et al.'s (ref. 1) sequence definition and terminology. Vertical thick-dashed lines delimit megasequence boundaries, vertical thin-dashed lines delimit supersequence set boundaries. Supersequence sets are as follows. T-A and T-B: Tejas A and B. Z-C, Z-B and Z-A: Zini C, B and A. A-UB, A-UA and A-L: upper Absaroka B, upper Absaroka A, and lower Absaroka. KA-U and KA-L: upper Kaskaskia and lower Kaskaskia. TR-U and TR-L: upper Trippecanoe and lower Trippecanoe. SA-U, SA-M and SA-L: upper Sauk, middle Sauk, and lower Sauk. **(b)** Gaussian bandpass filter output (0.028 ± 0.012 my$^{-1}$ frequency cutoff) to recover the ~36 Myr cycle. **(c)** Detrended eustatic curve: residuals of the 25% weighted average in 'a'; C1, C2, M1-M5, and P1-P7 are the ~36 Myr cycles (roughly 'C' for Cenozoic, 'M' for Mesozoic, and 'P' for Paleozoic), question marks indicate that ~36 Myr cycle boundaries are uncertain. Gaussian bandpass filter output (0.011 ±0.005 Myr$^{-1}$ frequency cutoff) to recover the ~91 Myr cycle (in blue), the ~36 Myr cycle bundlings are also shown with arcs. **(d)** Geologic time scale 2016, GTS2016 (ref. 5), Period: white box, Quaternary, N.: Neogene, Paleog: Paleogene, Carbonif: Carboniferous, Silur: Silurian, Ordovic: Ordovician.

## II. Very long-term eustatic cycles vs. other geologic process during the Phanerozoic Eon

Statistical significance and possible origin of multi-Myr cycles in geological records (e.g., biomass extinctions, climate change, Earth's interior dynamics, eustatic changes, etc) have been considered by several researchers since the 1980's (ref. 6-24, and many others).

In particular, a series of geologic events have been suggested to be related with a dominant ~30 ±5 Myr periodicity (e.g., ref. 10-13). The regularity, statistical significance, and origin of this dominant ~30 ±5 Myr periodicity have, however, been the subject of a long debate (e.g., ref. 6-8, 20, 25).

One hypothesis to explain biotic extinctions relates the cyclic variations to changes in the flux of the cosmic rays (CR) or of the Oort cloud comet, both in response to the Sun's oscillation about the galactic midplane (e.g., ref. 22, 24, 26).

In this study, we show that the most recent and revised Phanerozoic eustatic curve documents a prominent and continuous ~36 Myr cyclicity superimposed on Cenozoic-Mesozoic and Paleozoic megacycles (~250-Myr-long) (Fig. S4). These cyclicities were also suggested from several geological proxies (production rate on oceanic crust, temperature, biodiversity) (Fig. S4). Moreover, the most constrained deep-sea Cenozoic $\delta^{18}$O data show the ~36 Myr cycle, matching well the eustatic cycle (Fig. 3 from the main text). The ~35 Myr Cenozoic $\delta^{18}$O cycle was even previously recognized by Kaiho and Saito (ref. 13).

Our study outlines, for the first time, highly significant ~36 and ~250 Myr periodicities in a relatively well-constrained Phanerozoic sea-level record. Such periodicities are *à priori* astronomically predicated. The ~36 Myr period is equivalent to half-period of solar-system vertical motion. The ~250 Myr period could correspond to a period of radial solar-system motion (see SM2). This intriguing correspondence may hint at a possible connection between astronomy and geology. We then build a CR model that takes into account vertical and radial motions of the solar system in the galaxy, and may explain longer-term glacioeustatically driven SL changes (see SM2).

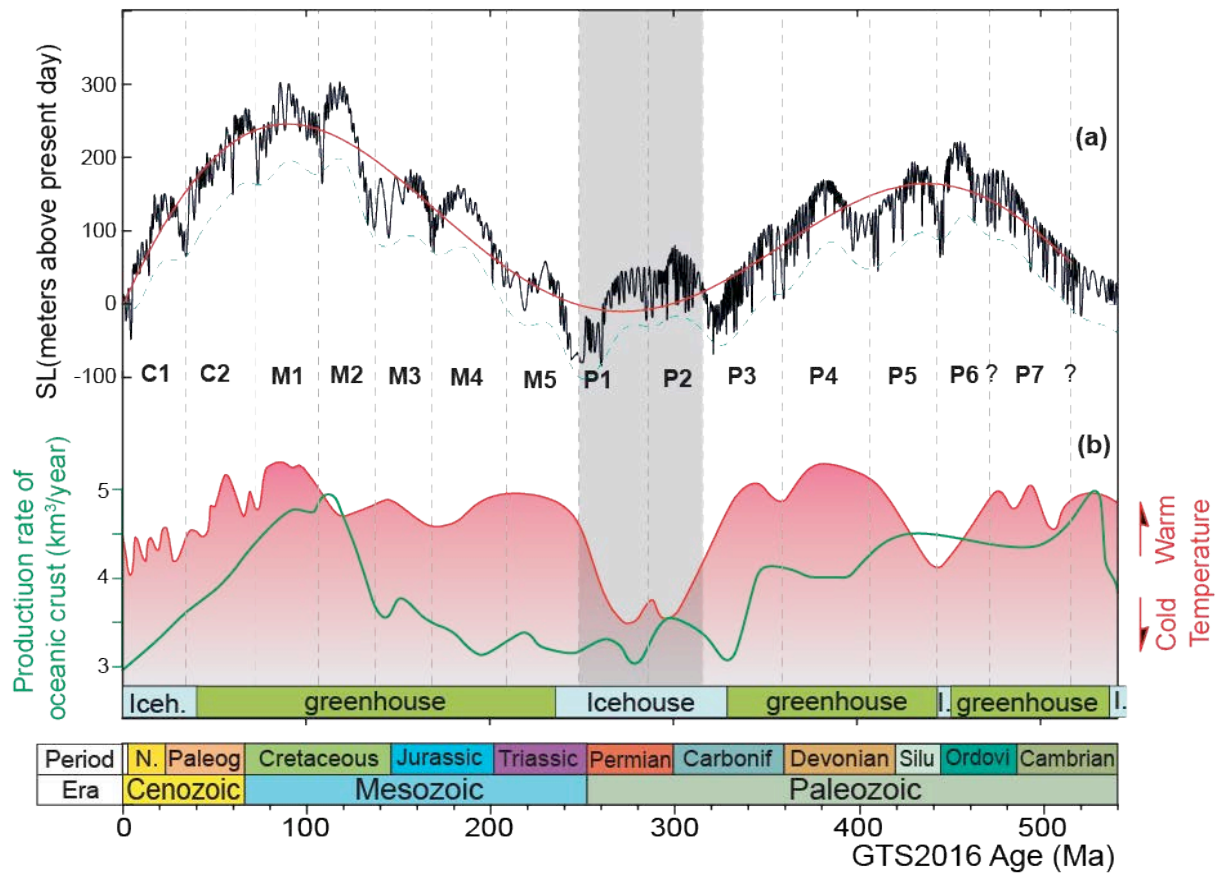

**Figure S4:** Comparison of the Phanerozoic eustatic variations with other geologic events. **(a)** Original Phanerozoic eustatic curve (ref. 1, 3). Cenozoic-Mesozoic and Paleozoic megacycles are fitted with six-order polynomial method (green curve). Lowpass Gaussian filtering (blue curve) highlights both ~36 Myr cycles (labelled as in Fig. 1b in the main text) and megacycles. **(b)** Oceanic crust production rates (green curve) are from Stanley (ref. 29), temperature variations (red curve) are from Frakes et al. (ref. 30), and climate model (icehouse vs greenhouse periods) is from Frakes et al. (ref. 30) where Cenozoic icehouse period is updated according to Zachos et al. (ref. 27, 28), grey-shaded 'P1' and 'P2' indicate a possible node (minimum) in the megacycle variations.

### III. Supporting information on Phanerozoic δ¹⁸O data

A recent study of Shaviv et al. (2014) (ref. 31) have used $\delta^{18}O$ data from specific fossils to conclude the persistence of the 32 Myr cyclicity throughout the Phanerozoic eon. In particular, they have used two $\delta^{18}O$ compilations: (1) $\delta^{18}O$ "ML200" compilation is considered as the "master" record used to highlight the 32 Myr cyclicity, and (2) $\delta^{18}O$ "ML175" compilation, which is close to ML200 but reduce jump between datasets). Both compilations do not include Cenozoic deep-sea $\delta^{18}O$ data, while those data show evidence of the 35 Myr cyclicity (Fig. S5). Here, we show that the temporally most constrained 0-65 Ma interval (Cenozoic) from their both (raw) compilations ML200 and ML175 does not show obvious 32 Myr cycles (Fig. S5). Instead, we argued (as in the main text, see main Fig. 5) that deep-sea (benthic foraminifera) composite $\delta^{18}O$ curve (e.g., ref. 27, 28) detects faithfully the ~32 Myr cyclicity (Fig. S6), pointing to glacio-eustatically driven SL change.

Accordingly, we have compiled $\delta^{18}O$ data as follows. We focused on the past 202 Ma interval because, with the exception of the interval around 120 Ma, (i) it includes high resolution data, (ii) it represents less $\delta^{18}O$ amplitude fluctuations (less scatter), and (iii) it possesses less significant observational gaps. For the interval 0–112 Ma, we used only the high-resolution deep-sea (benthic foraminifera) data (ref. 27, 28, 32 and 33). For the interval 112–202 Ma, we used only $\delta^{18}O$ data from brachiopods and belemnites (ref. 34) to reduce differential effects from fossil groups on $\delta^{18}O$ values. The resulting (202-Myr-long) compiled $\delta^{18}O$ signal is then calibrated to the recent geologic time scale GTS2016 (Fig. S7).

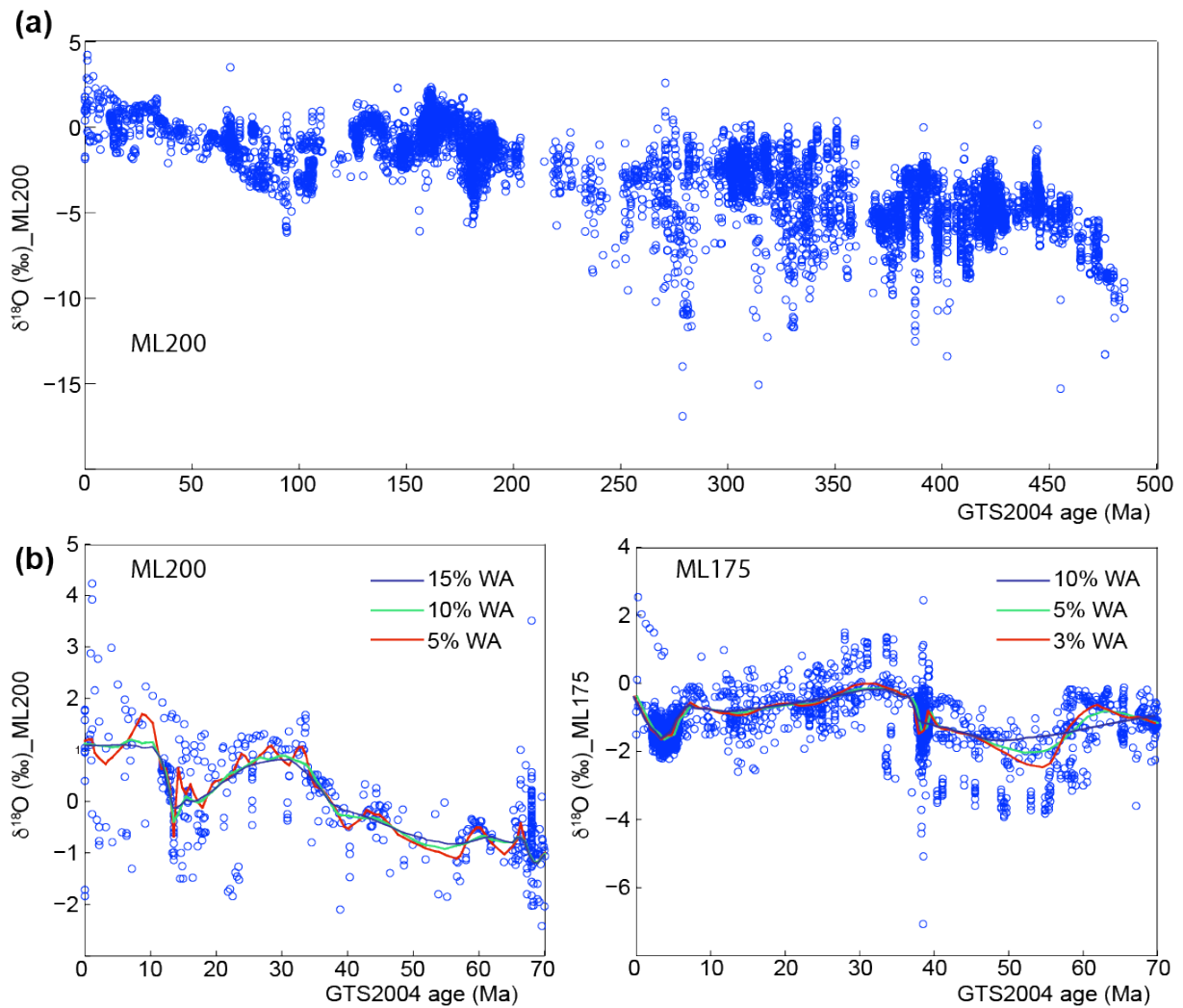

**Figure S5:** Phanerozoic $\delta^{18}O$ data (ref. 31). **(A)** $\delta^{18}O$ data for the interval 0-490 Ma. **(B)** $\delta^{18}O$ data for the interval 0-70 Ma from ML200 compilation (left panel), and ML75 compilation (right panel). Weighted Averages (WA) are applied using different smoothing factors (15%WA, 10%WA, 5%WA and 3%WA). Note that there is no clear cyclicity at 32 Myr band neither in ML200 nor in ML175 compilations in the Cenozoic interval.

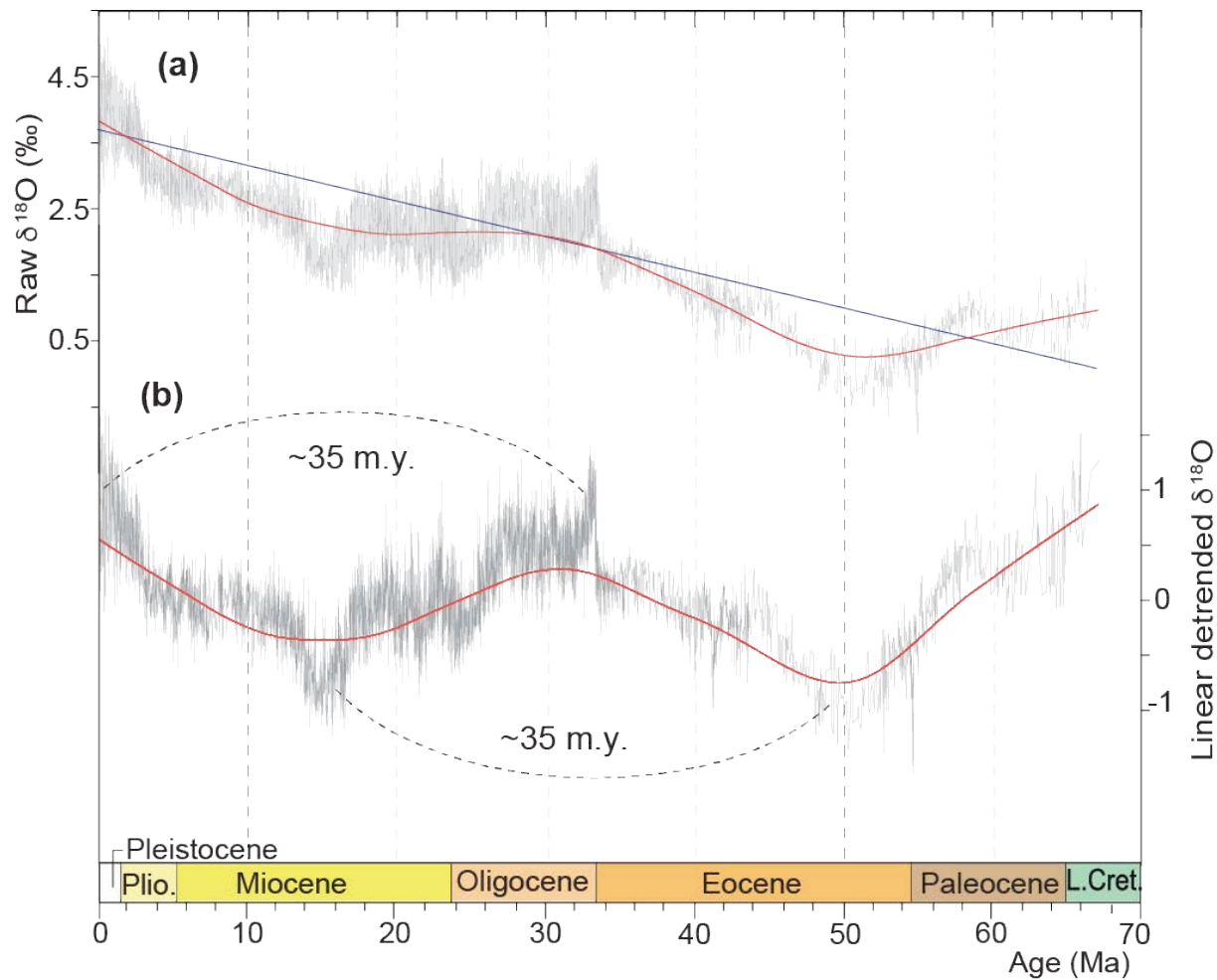

**Figure S6:** Strong ~35 Myr cyclicity in the Cenozoic $\delta^{18}O$ data. **(a)** Raw benthic foraminiferal oxygen isotopes $\delta^{18}O$ (ref. 27, 28), linear trend and a 25% weighted average of the series are also shown. **(b)** Linear-detrended $\delta^{18}O$, a 25% weighted average of the linear-detrended series and the strong ~35 Myr cycle are shown.

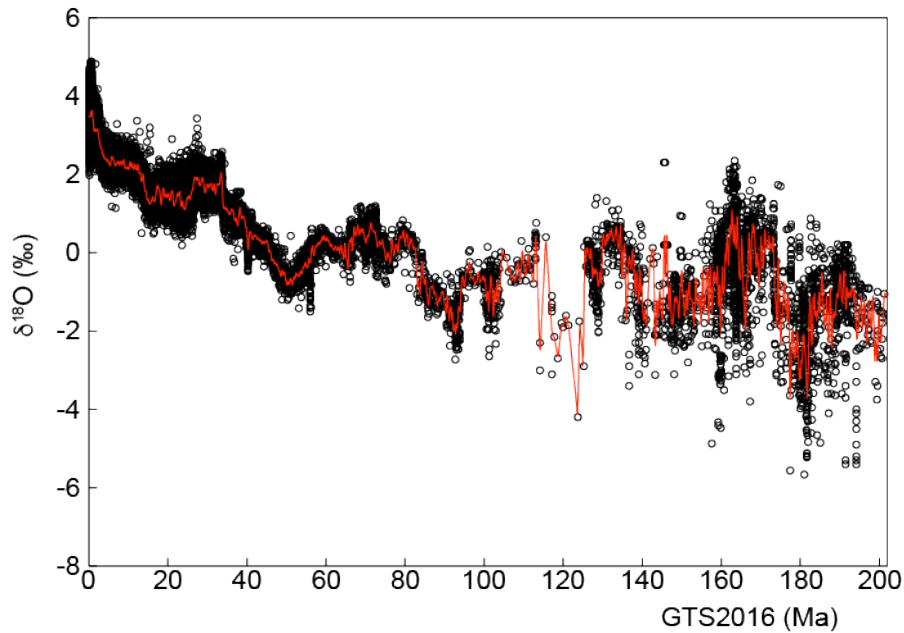

**Figure S7:** δ$^{18}$O data over the past 202 Ma (ref. 32,33,34) calibrated to the Geologic Time Scale 2016, GTS2016 (see text for details about compilation). The red curve is the fitted data using the least-square method.

## REFERENCES CITED

1. Haq BU, Hardenbol J, Vail PR (1987) Chronology of fluctuating sea levels since the Triassic. Science 235:1156–1167.
2. Boulila, S., Galbrun, B., Laskar, J., and Pälike, H., 2012, A ~9 myr cycle in Cenozoic δ$^{13}$C record and long-term orbital eccentricity modulation: Is there a link?: Earth Planet Sci. Lett., v. 317–318, p. 273–281.
3. Haq BU, Al-Qahtani AM (2005) Phanerozoic cycles of sea-level change on the Arabian Platform. GeoArabia 10:127–160.
4. Haq BU, Schutter SR (2008) A chronology of Phanerozoic Sea-Level Changes. Science 322:64–68.
5. Ogg, J.G., Ogg, G., and Gradstein, F.M. (2016) A Concise Geologic Time Scale 2016: Elsevier, store.elsevier.com/9780444637710, 234 p.
6. Raup DM, Sepkoski JJJr (1984) Periodicity of extinctions in the geologic past. Proc. Nat. Acad. Sci. USA 81:801-805.
7. Raup DM, Sepkoski JJJr (1988). Testing for periodicity of extinction. Science 241:94-96.
8. Rampino MR, Stothers RB (1984) Geological rhythms and cometary impacts. Science 226:142–143.
9. Rampino MR, Stothers RB (1988) Flood basalt volcanism during the past 250 million years. Science 241:663–668.
10. Rich JE, Johnson GL, Jones JE, Campsie J (1986) A significant correlation between fluctuations in seafloor spreading rates and evolutionary pulsations. Paleoceanography 1:85–95.
11. Negi JG, Tiwari RK, Rao KNN (1990) 'Clean' spectral analysis of long-term sea-level changes.

Terra Nova 2:138–141.

12. Rampino RM, Caldeira K (1993) Major episodes of geologic change: correlations, time structure and possible causes. Earth Planet Sci Lett 114:215–227.
13. Kaiho K, Saito S (1994) Oceanic crust production and climate during the last 100 Ma. Terra Nova 6:376–384.
14. Hallam A, Wignall PB (1999) Mass extinctions and sea-level changes. Earth Sci Rev 48:217–250.
15. Prokoph A, Rampino MR, Bilali H (2004) Periodic components in the diversity of calcareous plankton and geological events over the past 230 my. Palaeogr Palaeoclim Palaeoecol 207:105–125.
16. Prokoph A, Shields GA, Veizer J (2008) Compilation and time-series analysis of a marine δ18O, δ13C, 87Sr/86Sr and δ34S database through Earth history. Earth Sci Rev 87:113–133.
17. Rohde RA, Muller RA (2005) Cycles in fossil diversity. Nature 434:208–210.
18. Omerbashich M (2006) A Gauss-Vaníček Spectral Analysis of the Sepkoski Compendium : No New Life Cycles. Comput Sci Engineer 9:26-30.
19. Cornette JL (2007) Gauss-Vaníček and Fourier Transform Spectral Analyses of Marine Diversity. Comput Sci Engineer 9:62, doi: 10.1109/MCSE.2007.76
20. Lieberman BS, Melott AL (2007) Considering the Case for Biodiversity Cycles: Reexamining the Evidence for Periodicity in the Fossil Record. PLoS One 2(8):e759 doi:10.1371/journal.pone.0000759.
21. Alroy J, et al. (2008) Phanerozoic Trends in the Global Diversity of Marine Invertebrates. Science 312:97–100.
22. Melott AL (2008) Long-term cycles in the history of life: Periodic biodiversity in the Paleobiology Database. PloS One 3(12):e4044. doi:10.1371/journal.pone.0004044
23. Bailer-Jones CAL (2009) The evidence for and against astronomical impacts on climate change and mass extinctions: A review. Int J Astrob 8:213–239.
24. Melott AL, Bambach RK (2011) An ubiquitous ~62 my periodic fluctuation superimposed on general trends in fossil biodiversity: Part I, Documentation. Paleobiology 37:92-112.
25. Sepkoski JJJr (1989) Periodicity in extinction and the problem of catastrophism in the history of life. J Geol Soc London 146:7–19.
26. Matese JJ, Whitman PG, Innanen KA, Valtonen MJ (1995) Periodic Modulation of the Oort Cloud Comet Flux by the Adiabatically Changing Galactic Tide. Icarus 116:255–268.
27. Zachos JC, Dickens GR, Zeebe RE (2008) An early Cenozoic perspective on greenhouse warming and carbon-cycle dynamics. Nature 451:279–283.
28. Zachos JC, Pagani M, Sloan L, Thomas E, Billups K (2001) Trends, Rhythms, Aberrations in Global Climate 65 Ma to Present. Science 292:686–693.
29. Stanley SM (1999) Earth System Histrory. WH Freeman and Company, New York, 615 p.
30. Frakes LA, Francis JE, Syktus JI (1992) Climate Modes of the Phanerozoic: The History of the Earth's Climate Over the Past 600 Million Years. Cambridge University Press 274 p.
31. Shaviv, N.J., Prokoph, A., and Veizer, J., 2014, Is the Solar System's Galactic Motion Imprinted in the Phanerozoic Climate?: Scientific Reports, v. 4, no. 6150, doi:10.1038/srep06150.

32. Cramer, B.S., Toggweiler, J.R., Wright, J.D., Katz, M.E., and Miller, K.G., 2009, Ocean overturning since the Late Cretaceous: Inferences from a new benthic foraminiferal isotope compilation. Paleoceanography, v. 24, PA4216, doi: 10.1029/2008PA001683.
33. Friedrich, O., Norris, R.D., and Erbacher, J., 2012, Evolution of middle to Late Cretaceous oceans—A 55 m.y. record of Earth's temperature and carbon cycle. Geology, v. 40, p. 107–110.
34. Veizer, J., and Prokoph, A., 2015, Temperatures and oxygen isotopic composition of Phanerozoic oceans: Earth-Science Reviews, v. 146, P. 92–104.

**Supplementary material SM2**

Supplementary Information (2)
Galactic trajectories and cosmic rays

# 1 Models

The precise form of the galactic potential is still largely unknown. A variety of models exists in the literature (see Binney and Tremaine, 2008), but despite improvements in the constraints resulting from Hipparchos data, strong constraints are still missing (Dehnen and Binney, 1998). Because of the lack of constraints, we choose to use a model for the galactic potential that is as simple as possible, derived from the one of (Paczynski, 1990). This model comprises a bulge, a disk and a halo.

The bulge and disk of mass $M_1$ and $M_2$ are given by Miyamoto and Nagai potential $\Phi_1$ and $\Phi_2$ of the form (Miyamoto and Nagai, 1975)

$$\Phi_i(r,z) = -\frac{GM_i}{(r^2 + \left[a_i + (z^2 + b_i^2)^{1/2}\right]^2)^{1/2}}$$

with associated density

$$\rho_i(r,z) = \frac{b_i^2 M}{4\pi} \frac{a_i r^2 + (a_i + 3(z^2 + b_i^2)^{1/2})(a_i + (z^2 + b_i^2)^{1/2})^2}{(r^2 + (a_i + (z^2 + b_i^2)^{1/2})^2)^{5/2}(z^2 + b_i^2)^{3/2}} \ .$$

The potential of the Halo component is (Paczynski, 1990)

$$\Phi_h = \frac{GM_h}{h}\left[\frac{1}{2}\log\left(1 + \frac{R^2}{h^2}\right) + \frac{h}{R}\arctan(\frac{R}{h})\right] \ .$$

with $R = \sqrt{r^2 + z^2}$ and associated density

$$\rho_h = \frac{M_h}{4\pi h(R^2 + h^2)} \ .$$

The total potential is $\Phi = \Phi_1 + \Phi_2 + \Phi_h$. The equations of motion will be

$$\ddot{r} - r\dot{\phi}^2 = -\frac{\partial \Phi}{\partial r}$$

$$\frac{d}{dt}(r^2 \dot{\phi}) = -\frac{\partial \Phi}{\partial \phi} = 0$$

$$\ddot{z} = -\frac{\partial \Phi}{\partial z}$$

Due to rotational symmetry, we have $r^2\dot\phi = C$ constant (angular momentum conservation). We have also the conservation of the total energy per unit mass ($dH/dt = 0$) with

$$H = \frac{1}{2}(\dot r^2 + r^2\dot\phi^2 + \dot z^2) + \Phi(r, z) \ .$$

The system can thus be reduced to a system of order 4, but we prefer to use a phase space of dimension 5 with the integral of the energy as a verification for the accuracy of the integration. We use the set of variables $(r, \phi, z, p_r, p_z)$ with $p_r = \dot r$, $p_z = \dot z$ and the system of equations of first order

$$\begin{aligned}
\dot r &= p_r \\
\dot\phi &= \frac{C}{r^2} \\
\dot z &= p_z \\
\dot p_r &= -\frac{\partial\Phi}{\partial r} + \frac{C^2}{r^3} \\
\dot p_z &= -\frac{\partial\Phi}{\partial z}
\end{aligned} \quad (1)$$

## 2 Parameters and initial conditions

The initial conditions for the Sun (Binney et al., 1997; Reed, 2006; Schönrich et al., 2010) are given in Table 1. The constraint on the longitudinal velocity of the Sun is (Reid and

Table 1: Initial conditions.

| | |
|---|---|
| $r_0$ | 8300 pc $\pm 300$ |
| $z_0$ | 14 pc $\pm 4$ |
| $\phi_0$ | 0 |
| $\dot r_0$ | $-11$ km/s $\pm 1$ |
| $\dot z_0$ | 7 km/s $\pm 0.5$ |

Brunthaler, 2004).

$$v_0 = r_0\dot\phi_0 = 236 \pm 15 km/s$$

which provides the value of $C = r_0^2\dot\phi_0 = r_0 v_0$. The gravitational constant $G$, expressed in pc and Myr is $G = 0.0044755$ (equivalent to $4.302 \times 10^{-3}$ pc $M_\odot^{-1}$ (km/s)$^2$ with 1 pc/Myr=0.98 km/s).

### 2.1 Parameters

For our model, we have chosen parameters that are slightly different from the ones of (Paczynski, 1990) (Table 2). To determined these parameters, we iterated some fitting process in order to retrieve as close as possible the features observed in the geological data, keeping the initial conditions that are provided by stellar data (Table 1). This model can be though as a galactic model that is both fitted to stellar data and to geological data. It departs slightly from conventionally adopted galactic models as (Paczynski, 1990), but is still fully compatible with the up to date observational data (Table 1) (Schönrich et al., 2010).

Table 2: Parameters of the potential. P90 design the parameters from the model of (Paczynski, 1990)

|  | P90 | This study |
|---|---|---|
| $M_1$ | 1.12 E10 $M_\odot$ | 2.3 E10 $M_\odot$ |
| $a_1$ | 0 | 0 |
| $b_1$ | 277 pc | 277 pc |
| $M_2$ | 8.07 E10 $M_\odot$ | 9.5E10 $M_\odot$ |
| $a_2$ | 3700 pc | 4600 pc |
| $b_2$ | 200 pc | 193 pc |
| $M_h$ | 5. E10 $M_\odot$ | 5. E10 $M_\odot$ |
| $h$ | 6000 pc | 24 600 pc |

With this set of parameters, we obtain a stellar density in the vicinity of the Sun of $0.213 M_\odot/\text{pc}^3$ which is slightly larger than the value $0.158 M_\odot/\text{pc}^3$ of (Paczynski, 1990).

## 2.2 Numerical integration

The equations of motions are integrated using a Runge-Kutta method of order 8/7 (Hairer et al., 1993). The maximum variation of the relative energy smaller than $3.5 \times 10^{-15}$ over 2 Gyr. The orbit present vertical and radial oscillations with respective periods 72 and 254 Myr (Figs.1,2).

# 3 Cosmic Rays

Primary cosmic rays, up to a few Tev are created by violent phenomena in the galaxy, as explosions of stars, that can be traced by supernovae remnants and pulsars (see Delahaye, 2010). The distribution of these potential sources of cosmic rays in the Milky Way can be modelized by an expression of the form

$$\rho(r, z) = \rho_0 r^a \exp\left(-\frac{r}{r_0}\right) \exp\left(-\frac{|z|}{z_0}\right) . \qquad (2)$$

Although this general expression is in agreement with the observational results, some large uncertainty remains for the determination of the parameters in this expression (see Delahaye, 2010). Here we are mostly interested in the qualitative aspect of this distribution, and we have chosen the set of parameters L04 from (Lorimer, 2004) with $a = 2.35$, $r_0 = 1528$ pc, and $z_0 = 100$ pc (as in (Delahaye, 2010)). The expression (2) provides the distribution of the cosmic rays sources, but we are interested in the distribution of the cosmic rays in the Milky Way, and more particularly in the vicinity of the Solar orbit. Ideally, we would need to make some model for cosmic rays propagation, which is an involved process which relies on many unknown parameters (see Delahaye, 2010). On the other hand, the resulting distribution of cosmic rays, as depicted for example by the figure 7.2 of (Delahaye, 2010) seems to rely on simple smooth functions, that could thus be obtain by relatively simple reasoning.

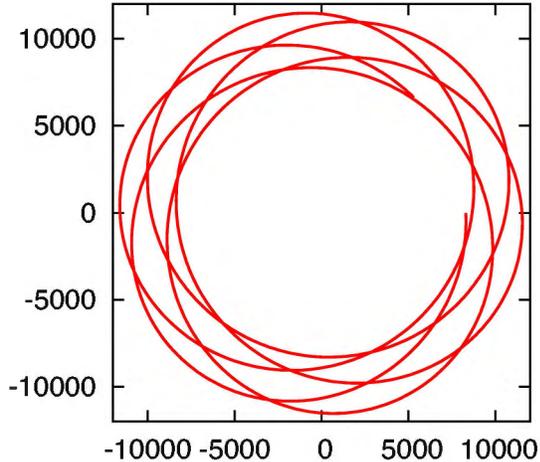

Figure 1: Projection of the orbit of the Sun in the galactic plane over 1.5 Gyr in the past. Units are in parsecs (pc).

## 3.1 Distribution of cosmic rays in the Milky Way

In order to have a qualitative model for the distribution of cosmic rays, we simply assume that the cosmic rays propagation decreases as the inverse of the square of the distance from the source, as for any beam isotrope propagation. When limited to the galactic plane ($z = 0$), the flux at distance $x$ from the galactic center will then be

$$\gamma_r(x) = \int_0^{+\infty} \rho_r(r) \frac{1}{|x-z|^c} dr \qquad (3)$$

where $\rho_r(r) = \rho(r,0)$ in (2) and $c = 2$. This function is then normalized such that $\gamma_r(r_0) = 1$. Using the L04 set of parameters for the sources, we obtain the distribution of cosmic rays given in Figure 3 (in red). In a similar way, the distribution in $z$ is obtained as

$$\gamma_z(x) = \int_0^{+\infty} \rho_z(z) \frac{1}{|x-z|^c} dz \qquad (4)$$

where $\rho_z(z) = \rho(r_0, z)$ in (2) and $c = 2$. The resulting distribution is given in Figure 4 (in red). As the hypothesis of isotropic distribution of the beam is an extreme case, we have also considered the case when the beam is confined in the galactic disk. In this case, the law of propagation is given by (3, 4), but with $c = 1$. The corresponding curves are displayed in blue in figures 3, 4.

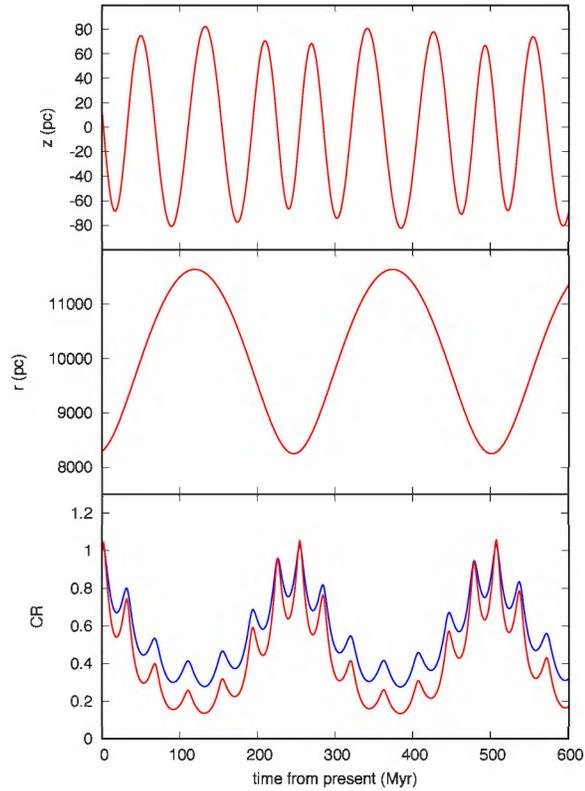

Figure 2: Top : variation of the vertical component ($z$ in pc) of the Sun with respect to the galactic plane over 600 Myr in the past. Middle : Radial distance ($r$ in pc) from the galactic center. Bottom : estimate of the variation of the cosmic ray flux on Earth resulting from the trajectory of the Sun in the Galaxy over 600 Myr in the past. In red with a decay in $1/d^2$, in blue with a decay in $1/d$.

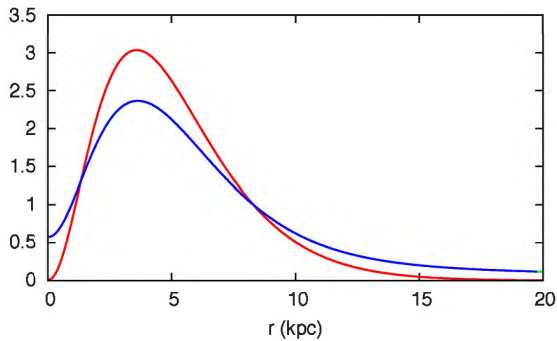

Figure 3: Distribution of cosmic rays as a function of $r$ (in pc) obtained with the L04 distribution of sources and the model of propagation (3). The distribution is normalized in such way that it is equal to unity for the present location of the Sun ($r_0 = 8300$ pc). Red : decay in $1/d^2$. Blue : decay in $1/d$.

## 3.2 Adjusted laws

To make it more easy to handle, models are now fitted to the previous results in order to obtain analytical expressions for the rate of cosmic rays in the galaxy. We thus fit simple models to the previous results. For the evaluation in the $z$ direction, we limit ourselves to 100 pc, which is larger than the excursion of the Sun. We can thus use a simple model of the form

$$\tilde{\gamma}_z(z) = \exp\left(a_1 |z| + a_2 |z|^2 + a_3 |z|^3 + a_4 |z|^4 + a_5 |z|^5\right) \tag{5}$$

that fits very well the numerical data obtained by solving the propagation law (4) (see Fig. 5). The coefficients $a_i$ for the $1/d$ and $1/d^2$ models are provides in Table 3.

Table 3: Coefficient of the approximated formula (5) for the $1/d$ and $1/d^2$ models

|       | $1/d$         | $1/d^2$       |
|-------|---------------|---------------|
| $a_1$ | $+1.3330E-04$ | $+4.8829E-05$ |
| $a_2$ | $-1.7432E-04$ | $-3.2842E-04$ |
| $a_3$ | $+2.6428E-06$ | $+5.8540E-06$ |
| $a_4$ | $-2.0977E-08$ | $-5.0908E-08$ |
| $a_5$ | $+6.6520E-11$ | $+1.7015E-10$ |

For the distribution in $r$, as the resulting law is more complex when the decay is in $1/d$, we choose to approximate the distribution in the $[5 : 15]$ kpc range with a polynomial of degree 3, with $\tilde{\gamma}_r(r_0) = 1$, as this is all we need for the evaluation of cosmic rays intensity in the vicinity of the Sun orbit. We thus obtain

$$\tilde{\gamma}_r(r) = 1 - 2.60 \times 10^{-04}(r - r_0) + 2.68 \times 10^{-08}(r - r_0)^2 - 7.94 \times 10^{-13}(r - r_0)^3 \quad (6)$$

for the decay law in $1/d$, and

$$\tilde{\gamma}_r(r) = 1 - 3.54 \times 10^{-04}(r - r_0) + 4.41 \times 10^{-08}(r - r_0)^2 - 1.80 \times 10^{-12}(r - r_0)^3 \ , \quad (7)$$

where $r$ is in pc, for the decay in $1/d^2$ (Fig.6).

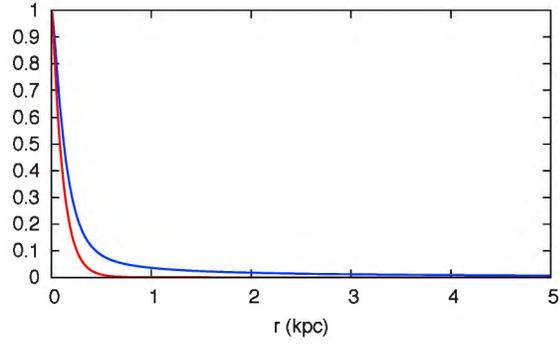

Figure 4: Distribution of cosmic rays as a function of $z$ (in pc) obtained with the L04 distribution of sources and the model of propagation (3). The distribution is normalized in such way that it is equal to unity for the present location of the Sun ($z_0 = 14$ pc). Red : decay in $1/d^2$. Blue : decay in $1/d$.

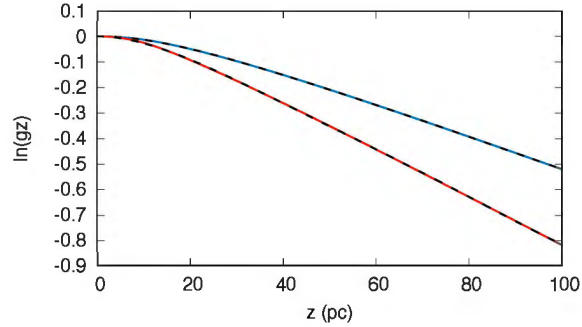

Figure 5: Distribution of cosmic rays as a function of $z$ (in pc) obtained with the L04 distribution of sources and the model of propagation (3). In red : decay in $1/d^2$. In blue : decay in $1/d$. In both cases, the solid line is the numerical computation from the model $\gamma_z(z)$ 4, while the dotted line is the fitted model $\tilde{\gamma}_z(z)$ (4).

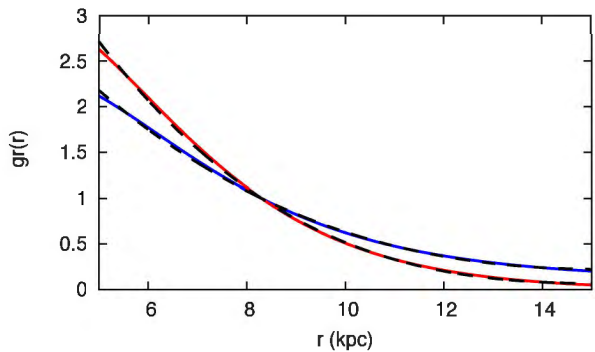

Figure 6: Distribution of cosmic rays as a function of $r$ (in kpc) obtained with the L04 distribution of sources and the model of propagation (3). After polynomial fit of degree 3 $\tilde{\gamma}_r(r_0) = 1$. Red : decay in $1/d^2$. Blue : decay in $1/d$. In both cases, the solid line is the numerical computation from the models (3,4), while the black dotted line is the polynomial fit $\tilde{\gamma}_r(r)$ (6,7).

It should be noted that both vertical and radial distributions are is good agreement with the latest observational results from (Abdo et al., 2008; Ackermann et al., 2012; Abramowski et al., 2014; Bartoli et al., 2015; Chen et al., 2015; Acero et al., 2016).

The evolution of the orbit of the Sun in the Galaxy is then obtained by integrating the equations of motion (1), and the cosmic rays flux on the Solar System through time is obtained using the analytical approximate formula $\tilde{\gamma}_r(r)$ and $\tilde{\gamma}_z(z)$ with a total flux

$$\tilde{\gamma}(r,z) = \tilde{\gamma}_r(r)\tilde{\gamma}_z(z) \ . \quad (8)$$

The computed flux is given in Fig.2 (bottom) for the decay law in $1/d^2$ (in red) and the decay law in $1/d$ (in blue). It can be seen that the results do not differ much when we change the law of propagation of the cosmic rays in the Galaxy. Either curve can thus be used. We provide also in Fig.7 the evolution of the cosmic rays flux when only the vertical component of the motion (in $z$) is considered.

## 3.3 Discussion

In this study, we have used a potential for the Milky way with a rotational symmetry, as well as the distribution of cosmic rays sources. We have shown that a $\sim 250$ Myr periodicity can be retrieved in the rate of cosmic rays on the Solar System by considering the radial excursion of the Sun. It should be noted that although the 72 Myr vertical period is very robust, the obtention of the 250 Myr period through radial variation of the Sun orbit requires some fine tuning that is made possible by the uncertainty that remains at present on the knowledge of the vicinity of the Sun and structure of the Milky way. With the future results of the GAIA astrometric mission, this uncertainty should be highly reduced.

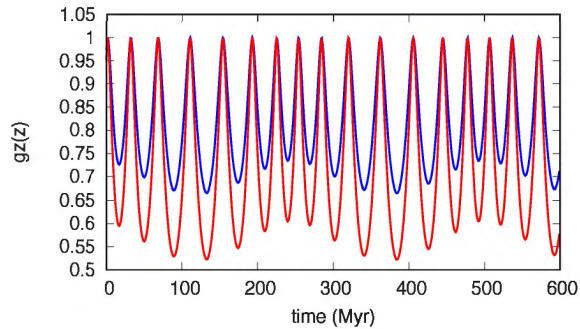

Figure 7: Distribution of cosmic rays as a function of time (in Myr in the past) obtained with the L04 distribution of sources and the model of propagation (3). only the vertical component $\tilde{\gamma}_z(z)$ of the motion is considered.

If the derivation of the 250 Myr radial period is still possible, the present study would then provide an additional important constraint on the structure of the Milky way. If on the contrary the new Gaia data rule out the possibility of a 250 Myr radial period, one would have to search again for a possible explanation of the observation of this signature in the geological record as the possible time scale for the passage of the Sun through the spiral arms of the galaxy (Shaviv, 2002; Gies and Helsel, 2005; Svensmark, 2006), assuming that Gaia data will also constraint much better the structure of the spirals arms in the galaxy.

# References

A. A. Abdo, et al. A Measurement of the Spatial Distribution of Diffuse TeV Gamma-Ray Emission from the Galactic Plane with Milagro. The Astrophysical Journal, 688: 1078–1083, December 2008. ISSN 0004-637X. Doi: 10.1086/592213.

A. Abramowski, et al. Diffuse Galactic gamma-ray emission with H.E.S.S. Physical Review D, 90: 122007, December 2014. ISSN 0556-2821. doi: 10.1103/PhysRevD.90.122007.

F. Acero, et al. Development of the Model of Galactic Interstellar Emission for Standard Point-source Analysis of Fermi Large Area Telescope Data. The Astrophysical Journal Supplement Series, 223:26, April 2016. ISSN 0067-0049. doi: 10.3847/0067-0049/223/2/26.

M. Ackermann, et al. Fermi-LAT Observations of the Diffuse -Ray Emission: Implications for Cosmic Rays and the Interstellar Medium. The Astrophysical Journal, 750:3, May 2012. ISSN 0004-637X. doi: 10.1088/0004-637X/750/1/3.

B. Bartoli, et al. Study of the Diffuse Gamma-Ray Emission from the Galactic Plane with ARGO-YBJ. The Astrophysical Journal, 806:20, June 2015. ISSN 0004-637X. doi: 10.1088/0004-637X/806/1/20.

James Binney and Scott Tremaine. Galactic Dynamics: Second Edition. 2008.

James Binney, Ortwin Gerhard, and David Spergel. The photometric structure of the inner galaxy. Monthly Notices of the Royal Astronomical Society, 288:365–374, June 1997.

Ding Chen, Jing Huang, and Hong-Bo Jin. Spectra of Cosmic Ray Electrons and Diffuse Gamma Rays with the Constraints of AMS-02 and HESS Data. The Astrophysical Journal, 811:154, October 2015. ISSN 0004-637X. doi: 10.1088/0004-637X/811/2/154.

Walter Dehnen and James Binney. Mass models of the milky way. Monthly Notices of the Royal Astronomical Society, 294:429, March 1998.

T. Delahaye. Propagation of galactic cosmic rays and dark matter indirect detection. PhD thesis, LAPTH, Universit´e de Savoie, Universit´e de Savoie, July 2010.

D. R. Gies and J. W. Helsel. Ice age epochs and the sun's path through the galaxy. The Astrophysical Journal, 626:844–848, June 2005.

Ernst Hairer, Syvert Paul Nørsett, and Gerhard Wanner. Solving ordinary differential equations: Nonstiff problems. Springer, 1993. ISBN 9783540566700.

D. R. Lorimer. The galactic population and birth rate of radio pulsars. In F. Camilo and BM Gaenler, editors, Young Neutron Stars and their Environments, volume 218 of IAU Symposium, page 105, 2004.

Miyamoto and Nagai. Three-dimensional models fro the distribution of mass in galaxies. PASJ, 27: 533–543, 1975.

Bohdan Paczynski. A test of the galactic origin of gamma-ray bursts. The Astrophysical Journal, 348:485–494, January 1990.

B. Cameron Reed. The sun's displacement from the galactic plane from spectroscopic parallaxes of 2500 OB stars. Journal of the Royal Astronomical Society of Canada, 100:146, August 2006.

M. J. Reid and A. Brunthaler. The proper motion of sagittarius a*. II. the mass of sagittarius a*. The Astrophysical Journal, 616:872–884, December 2004.

Ralph Schönrich, James Binney, and Walter Dehnen. Local kinematics and the local standard of rest. Monthly Notices of the Royal Astronomical Society, 403:1829–1833, April 2010.

Nir J. Shaviv. Cosmic ray diffusion from the galactic spiral arms, iron meteorites, and a possible climatic connection. Physical Review Letters, 89:51102, 2002.

H. Svensmark. Imprint of galactic dynamics on earth's climate. Astronomische Nachrichten, 327: 866, November 2006.